\documentclass[a4paper]{article}
\usepackage{geometry}
\usepackage{amsmath}
\usepackage{amssymb}
\usepackage{indentfirst}
\usepackage{mathdots}
\usepackage{cite}
\usepackage[colorlinks, linkcolor=blue, citecolor=blue]{hyperref}
\usepackage{booktabs}
\usepackage{enumitem}
\usepackage{mathrsfs}
\usepackage{enumitem}
\usepackage{authblk}
\usepackage{amsfonts,ntheorem}
\usepackage{float}
\usepackage{setspace}
\usepackage{color}

\makeatletter

\newcommand{\Rmnum}[1]{\expandafter\@slowromancap\romannumeral #1@}
\makeatother

\newtheorem{theorem}{Theorem}
\newtheorem{lemma}{Lemma}

\newtheorem{question}{Question}

{ 
\theoremstyle{plain}
\theorembodyfont{\normalfont} 

\newtheorem{remark}{Remark}

}

\newenvironment{proof}{\noindent{\textbf{\emph{Proof.}}}}

\begin {document}
\title{Galois hulls of MDS codes and their quantum error correction}

\author{Meng Cao \\{E-mail address: caom17@mails.tsinghua.edu.cn }}
\affil{Department of Mathematical Sciences, Tsinghua University, Beijing 100084, PR China }

\date{}

\maketitle

\begin{abstract}
The hull of linear codes plays an important role in quantum information and coding theory.
In the present paper, by investigating the Galois hulls of generalized Reed-Solomon (GRS) codes and extended GRS codes over the finite field $\mathbb{F}_{q}$, we give several new families of MDS codes with Galois hulls of arbitrary dimensions that are not obtained before. Some of them generalize the ones in the literature \cite{Fang2019Euclidean}. As a consequence, using these MDS codes with Galois hulls of arbitrary dimensions, we construct nine new families of MDS entanglement-assisted quantum error-correcting codes (EAQECCs) with flexible parameters.
\end{abstract}




\section{Introduction}
Let $\mathbb{F}_{q}$ denote the finite field with $q$ elements, where $q=p^{e}$ is a prime power. In \cite{Assmus1993Designs}, the hull of a linear code
$\mathcal{C}$ is defined as $Hull(\mathcal{C})=\mathcal{C}\bigcap \mathcal{C}^{\bot}$, where $\mathcal{C}^{\bot}$ is the Euclidean dual code of $\mathcal{C}$. When $Hull(\mathcal{C})=0$, the code $\mathcal{C}$ is just the linear complementary dual (abbreviated to LCD) code, which was introduced
by Massey \cite{Massey1992Linear}. He provided the algebraic characterization of this code and also proved that the asymptotically good LCD codes exist.
Afterwards, a great deal of research on LCD codes have been done by many scholars,
see \cite{Carlet2018Linear,Carlet2018Euclidean,Carlet2018New,Galindo2018New,Liu2015LCD,Jin2017Construction,Sok2020On}.
Moreover, the LCD code has many applications in other fields. For instance, Bringer et al. \cite{Bringer2014Orthogonal} and Carlet and Guilley \cite{Carlet2014Complementary} explored the problem of binary LCDs against side-channel attacks (SCAs) and fault injection attacks (FIAs).

Quantum error-correcting codes are essential to quantum computation and quantum
communication. After the great work in \cite{Calderbank1998Quantum,Shor1995Scheme,Steane1996Multiple}, the theory of quantum codes
has achieved great development these years. As we know, the construction of quantum
codes with good parameters is an important and difficult problem in quantum error-correction.
Fortunately, as a class of quantum error-correcting codes, stabilizer codes can be acquired
by the so-called CSS construction.
By means of this construction, an $[[n,2k-n,\geq d]]_{q}$ stabilizer code can be yielded if there exists an Euclidean dual-containing $[n,k,d]_{q}$ code.
However, in the above process, the classical linear code is required to be Euclidean dual-containing, which limits its applications to some extent.
To avoid this problem, a new concept called entanglement-assisted quantum
error-correcting codes (abbreviated to EAQECCs) was introduced by Brun et al. \cite{Brun2006Correcting}, which contain the
standard stabilizer codes as a special case. By utilizing pre-shared entanglement between the sender
and receiver, they showed that one can construct EAQECCs through the classical linear codes without the self-orthogonality restriction.
However, it turns out determining the
number of shared pairs that required to construct an EAQECC becomes a difficult thing.
In \cite{Guenda2018Constructions}, Guenda et al. gave the result that this number can be obtained from the
dimension of the hull of classical linear code.
Recently, Luo et al. \cite{Luo2019MDS} constructed several families of MDS EAQECCs by investigating the generalized Reed-Solomon (GRS)
codes and extended GRS codes with Euclidean hulls of arbitrary dimensions.
Afterwards, Fang et al. \cite{Fang2019Euclidean} studied the GRS
codes and extended GRS codes with Euclidean and Hermitian hulls of arbitrary dimensions. As a result, they obtained
several new classes of MDS EAQECCs with flexible parameters. These tell us that the study of the hull of linear codes, especially the MDS codes, is of great significance.

In \cite{Fan2017Galois}, Fan and Zhang generalized the Euclidean inner product and Hermitian inner product to a new notion called $l$-Galois form,
where $l$ is an integer with $0\leq l\leq e-1$. Therefore, we may desire to know how to deal with the construction of the GRS
codes and extended GRS codes with Galois hulls of arbitrary dimensions. In this paper, we will devote ourselves to
construct such codes, and thus obtain some new families of MDS EAQECCs.
We make two remaks on our results. Firstly, we provide several new families
of MDS codes with Galois hulls of arbitrary dimensions that are not obtained before.
Some of these MDS codes are constructed by means of certain multiplicative subgroups of $\mathbb{F}_{q}^{\ast}$ and their cosets.
Specially, some of them generalize the ones in \cite{Fang2019Euclidean}.
Secondly, through these MDS codes constructed in Section 3,
we obtain nine new classes of MDS EAQECCs with flexible parameters in Section 4. In particular, the required number of maximally entangled states can take
almost all possible values.

The remainder of this paper is structured as follows. Section 2 recalls some necessary concepts and properties about the Euclidean, Hermitian and Galois dual codes and (extended) GRS codes. In Section 3, we in turn present the constructions of GRS
codes and extended GRS codes with Galois hulls of arbitrary dimensions. In Section 4, applying the MDS codes obtained in Section 3, we obtain nine families of MDS EAQECCs, and we list some new parameters as an illustration. Section 5 gives a brief conclusion of this paper.

\section{Preliminaries}
Throughout this paper, $q$ is alway assumed to be a prime power $p^{e}$, where $p$ is a prime and $e$ is a positive integer.
Let $\mathbb{F}_{q}$ be the finite field with $q$ elements and let $\mathbb{F}_{q}^{\ast}$ denote the set of non-zero elements over $\mathbb{F}_{q}$.

A linear $[n,k,d]_{q}$ code $\mathcal{C}$ is a linear subspace of $\mathbb{F}_{q}^{n}$ with length $n$, dimension $k$ and minimum distance $d$. The minimum distance $d$ of a linear code $\mathcal{C}$ must satisfy the well-known Singleton bound $d\leq n+1-k$. We call a linear code $\mathcal{C}$ maximum distance separable (MDS) if its minimum distance achieves the above bound, i.e., $d=n+1-k$.

Given two vectors $\mathbf{x}=(x_{1},x_{2},\ldots,x_{n})$, $\mathbf{y}=(y_{1},y_{2},\ldots,y_{n})\in \mathbb{F}_{q}^{n}$, the Euclidean inner product of them is defined as
$$(\mathbf{x},\mathbf{y})_E=\sum_{i=1}^n x_i y_i.$$
For a linear code $\mathcal{C}$ of length $n$ over $\mathbb{F}_{q}$, the Euclidean dual code of $\mathcal{C}$ is defined as
$$ \mathcal{C}^{\perp_{E}}=\{\mathbf{x}\in\mathbb{F}_{q}^{n}|(\mathbf{x},\mathbf{y})_{E}=0 \ \mbox     {for  all}  \ \mathbf{y}\in \mathcal{C}\}.$$
Moreover, for two vectors $\mathbf{x}$, $\mathbf{y}\in \mathbb{F}_{q^{2}}^{n}$, their Hermitian inner product is defined as
$$(\mathbf{x},\mathbf{y})_H=\sum_{i=1}^n x_i y_{i}^{q}.$$
The Hermitian dual code of $\mathcal{C}$ is defined as
$$ \mathcal{C}^{\perp_{H}}=\{\mathbf{x}\in\mathbb{F}_{q^{2}}^{n}|(\mathbf{x},\mathbf{y})_{H}=0 \ \mbox     {for  all}  \ \mathbf{y}\in \mathcal{C}\}.$$
A linear code $\mathcal{C}$ is called Euclidean (or Hermitian) dual-containing if $\mathcal{C}^{\perp_{E}}\subseteq \mathcal{C}$ (or $C^{\perp _{H}}\subseteq \mathcal{C}$).

Apart from the Euclidean inner product and Hermitian inner product, Fan and Zhang \cite{Fan2017Galois} introduced a new notion called $l$-Galois form, which is defined as
$$(\mathbf{x},\mathbf{y})_l=\sum_{i=1}^n x_i y_{i}^{p^{l}},$$
where $l$ is an integer with $0\leq l\leq e-1$. One can see that $(\mathbf{x},\mathbf{y})_0$ is just the Euclidean inner product.
For an even integer $e$, $(\mathbf{x},\mathbf{y})_{\frac{e}{2}}$ is just the Hermitian inner product. Therefore, the concept
of $l$-Galois form generalizes the Euclidean inner product and Hermitian inner product. For any code $\mathcal{C}$ with
length $n$ over $\mathbb{F}_{q}$, the following code
$$ \mathcal{C}^{\perp_{l}}=\{\mathbf{x}\in\mathbb{F}_{q}^{n}|(\mathbf{y},\mathbf{x})_{l}=0 \ \mbox     {for  all}  \ \mathbf{y}\in \mathcal{C}\}$$
is called the $l$-Galois dual code of $\mathcal{C}$. If $\mathcal{C}\subseteq \mathcal{C}^{\perp_{l}}$, then we call
$\mathcal{C}$ a $l$-Galois self-orthogonal code. If $\mathcal{C}=\mathcal{C}^{\perp_{l}}$, then we call $\mathcal{C}$
a $l$-Galois self-dual code. Moreover, the $l$-Galois hull of $\mathcal{C}$ is defined as
$Hull_{l}(\mathcal{C})=\mathcal{C}\bigcap \mathcal{C}^{\perp_{l}}$.

For a vector $\mathbf{v}=(v_{1},v_{2},\ldots,v_{n})\in \mathbb{F}_{q}^{n}$, define ${\mathbf{v}}^{a}=(v_{1}^{a},v_{2}^{a},\ldots,v_{n}^{a})$ for
any integer $a$. Let $M$ be a subset of $\mathbb{F}_{q}^{n}$, then $M^{a}$ is defined as the set $\{{\mathbf{v}}^{a}|\mathbf{v}\in M\}$.

The following lemma characterizes the $l$-Galois dual code $\mathcal{C}^{\perp_{l}}$ of a linear code $\mathcal{C}$.
\begin{lemma}
(\cite{Liu2020New}) For an $[n,k,d]_{q}$ linear code $\mathcal{C}$, we have $\mathcal{C}^{\perp_{l}}=(\mathcal{C}^{p^{e-l}})^{\perp}$.
\end{lemma}

Take $\mathbf{a}=(a_{1},a_{2},\ldots,a_{n})$ with $a_{1},a_{2},\ldots,a_{n}$ being distinct elements in $\mathbb{F}_{q}$, and put $\mathbf{v}=(v_{1},v_{2},\ldots,v_{n})$ with $v_{1},v_{2},\ldots,v_{n}\in \mathbb{F}_{q}^{\ast}$. For $k\leq n\leq q$, the $k$-dimensional generalized Reed-Solomon (GRS) code with respect to $\mathbf{a}$ and $\mathbf{v}$ is defined as
\begin{align*}
GRS_{k}(\mathbf{a},\mathbf{v})=\{(v_{1}f(a_{1}),v_{2}f(a_{2}),\ldots,v_{n}f(a_{n}))|f(x)\in \mathbb{F}_{q}[x], \mbox{deg}(f(x))\leq k-1\}.
\end{align*}
It is an $[n,k,n-k+1]_{q}$ MDS code which has a generator matrix
$$G_{k}(\mathbf{a},\mathbf{v})=\left[
\begin{array}{cccc}
v_{1}& v_{2}&  \cdots &v_{n}\\[4pt]
v_{1}a_{1}&v_{2}a_{2}& \cdots &v_{n}a_{n}\\
\vdots&\vdots&\ddots&\vdots\\
v_{1}a_{1}^{k-1}& v_{2}a_{2}^{k-1}& \cdots &v_{n}a_{n}^{k-1}  \\
\end{array}\right].$$
Moreover, the extended GRS code with respect to $\mathbf{a}$ and $\mathbf{v}$ is defined as
$$GRS_{k}(\mathbf{a},\mathbf{v},\infty)=\{(v_{1}f(a_{1}),\ldots,v_{n}f(a_{n}),f_{k-1})|f(x)\in \mathbb{F}_{q}[x], \mbox{deg}(f(x))\leq k-1\},$$
where $f_{k-1}$ denotes the coefficient of $x^{k-1}$ in $f(x)$. It is not difficult to verify that $GRS_{k}(\mathbf{a},\mathbf{v},\infty)$ is an $[n+1,k,n-k+2]_{q}$ MDS code which has a generator matrix as follows:
$$G_{k}(\mathbf{a},\mathbf{v},\infty)=\left[
\begin{array}{ccccc}
v_{1}& v_{2}&  \cdots &v_{n}&0\\[4pt]
v_{1}a_{1}&v_{2}a_{2}& \cdots &v_{n}a_{n}&0\\
\vdots&\vdots&\ddots&\vdots&\vdots\\
v_{1}a_{1}^{k-2}& v_{2}a_{2}^{k-2}& \cdots &v_{n}a_{n}^{k-2}&0  \\[4pt]
v_{1}a_{1}^{k-1}& v_{2}a_{2}^{k-1}& \cdots &v_{n}a_{n}^{k-1}&1  \\
\end{array}\right].$$

Let $\mathbf{1}=(1,1,\ldots,1)$ be the all one vector. Along this paper, for each $i=1,2,\ldots,n$, we always denote
$$u_{i}=\prod_{1\leq j\leq n,j\neq i}(a_{i}-a_{j})^{-1},$$
which will be used frequently from now on.

Based on the above results, we obtain the following lemma.

\begin{lemma} \label{lemma1}

A codeword $\mathbf{c}=(v_{1}f(a_{1}),v_{2}f(a_{2}),\ldots,v_{n}f(a_{n}))$ of $GRS_{k}(\mathbf{a},\mathbf{v})$ is contained in $GRS_{k}(\mathbf{a},\mathbf{v})^{\bot_{l}}$ if and only if there exists a polynomial $g(x)$ with $\mbox{deg}(g(x))\leq n-k-1$, such that
$$(v_{1}^{p^{l}+1}f^{p^{l}}(a_{1}),v_{2}^{p^{l}+1}f^{p^{l}}(a_{2}),\ldots,v_{n}^{p^{l}+1}f^{p^{l}}(a_{n}))=(u_{1}g(a_{1}),u_{2}g(a_{2}),\ldots,u_{n}g(a_{n})).$$

\end{lemma}

\begin{proof}
It is not difficult to see that $G_{k}(\mathbf{a},\mathbf{v})=G_{k}(\mathbf{a},\mathbf{1})\cdot D$,
where $D=\mbox{diag}(v_{1},\ldots,v_{n})$.
Denote $\mathbf{u}=(u_{1},u_{2},\ldots,u_{n})$. By Lemma 1, for any $\mathbf{c}=(v_{1}f(a_{1}),v_{2}f(a_{2}),\ldots,v_{n}f(a_{n}))\in GRS_{k}(\mathbf{a},\mathbf{v})$, we have
\begin{align*}
\mathbf{c}\in GRS_{k}(\mathbf{a},\mathbf{v})^{\bot_{l}}&\Leftrightarrow \mathbf{c}\in (GRS_{k}(\mathbf{a},\mathbf{v})^{\bot})^{p^{e-l}}\Leftrightarrow
\mathbf{c}^{p^{l}}\in GRS_{k}(\mathbf{a},\mathbf{v})^{\bot}\\
&\Leftrightarrow G_{k}(\mathbf{a},\mathbf{v})(\mathbf{c}^{p^{l}})^T=0\Leftrightarrow G_{k}(\mathbf{a},\mathbf{1})D(\mathbf{c}^{p^{l}})^T=0\\
&\Leftrightarrow \mathbf{c}^{p^{l}}D\in GRS_{k}(\mathbf{a},\mathbf{1})^{\bot}=GRS_{n-k}(\mathbf{a},\mathbf{u}).
\end{align*}
The last equality follows by the result in \cite{Jin2014A}. $\hfill\square$
\end{proof}

\begin{remark} \label{Remark1}
Indeed, Lemma 2 is a generalization of \cite{Chen2018New} and \cite{Fang2018Two}, which consider the cases of Euclidean dual and Hermitian dual, respectively.
\end{remark}

Analogously, we obtain the following lemma for the case of extended GRS code $GRS_{k}(\mathbf{a},\mathbf{v},\infty)$.

\begin{lemma} \label{lemma3}

A codeword $\mathbf{c}=(v_{1}f(a_{1}),v_{2}f(a_{2}),\ldots,v_{n}f(a_{n}),f_{k-1})$ of $GRS_{k}(\mathbf{a},\mathbf{v},\infty)$ is contained in $GRS_{k}(\mathbf{a},\mathbf{v},\infty)^{\bot_{l}}$ if and only if there exists a polynomial $g(x)$ with $\mbox {deg}(g(x))\leq n-k$, such that
$$(v_{1}^{p^{l}+1}f^{p^{l}}(a_{1}),\ldots,v_{n}^{p^{l}+1}f^{p^{l}}(a_{n}),f_{k-1}^{p^{l}})=(u_{1}g(a_{1}),\ldots,u_{n}g(a_{n}),-g_{n-k}).$$

\end{lemma}

\begin{proof}
One can check that $G_{k}(\mathbf{a},\mathbf{v},\infty)=G_{k}(\mathbf{a},\mathbf{1},\infty)\cdot D'$,
where $D'=\mbox{diag}(v_{1},\ldots,v_{n},1)$. By Lemma 1,
for any $\mathbf{c}=(v_{1}f(a_{1}),v_{2}f(a_{2}),\ldots,v_{n}f(a_{n}),f_{k-1})$ of $GRS_{k}(\mathbf{a},\mathbf{v},\infty)$, we have
\begin{align*}
\mathbf{c}\in GRS_{k}(\mathbf{a},\mathbf{v},\infty)^{\bot_{l}}&\Leftrightarrow \mathbf{c}\in (GRS_{k}(\mathbf{a},\mathbf{v},\infty)^{\bot})^{p^{e-l}}\Leftrightarrow
\mathbf{c}^{p^{l}}\in GRS_{k}(\mathbf{a},\mathbf{v},\infty)^{\bot}\\
&\Leftrightarrow G_{k}(\mathbf{a},\mathbf{v},\infty)(\mathbf{c}^{p^{l}})^T=0\Leftrightarrow G_{k}(\mathbf{a},\mathbf{1},\infty)D'(\mathbf{c}^{p^{l}})^T=0\\
&\Leftrightarrow \mathbf{c}^{p^{l}}D'\in GRS_{k}(\mathbf{a},\mathbf{1},\infty)^{\bot}.
\end{align*}
Applying \cite[Lemma 5]{Fang2018Two}, one can complete the proof. $\hfill\square$
\end{proof}

\section{Construction of MDS codes with $l$-Galois hulls of arbitrary dimensions}
In this section, we will construct several families of MDS codes with $l$-Galois hulls of arbitrary dimensions.

\begin{theorem} \label{Theorem1}

For any $0\leq h\leq k$, there exists an $[n,k]_{q}$ MDS code $\mathcal{C}$ with $h$-dimensional $l$-Galois hull if one of the following conditions holds:

(1) $n\mid (q-1)$, $1\leq k\leq \lfloor\frac{p^{l}+n-1}{p^{l}+1}\rfloor$;

(2) $n\mid (p^{l}-1)$, $1\leq k\leq \lfloor\frac{n}{2}\rfloor$.

\end{theorem}

\begin{proof}
(1) For $n\mid (q-1)$, let $\alpha$ be a primitive $n$th root of unity in $\mathbb{F}_{q}$. Set $z:=k-h$.
Take $\mathbf{a}=(\alpha^{0},\alpha^{1},\ldots,\alpha^{n-1})$ and
$\mathbf{v}=(1,\ldots,1,v_{n-z+2},\ldots,v_{n})$ with $v_{j}\neq 0$ and $v_{j}^{p^{l}+1}\neq 1$ for $j=n-z+2,\ldots,n$.
We have
\begin{equation}
\begin{split}
u_{i}&=\prod_{1\leq j\leq n,j\neq i}(a_{i}-a_{j})^{-1}=\prod_{1\leq j\leq n,j\neq i}(\alpha^{i-1}-\alpha^{j-1})^{-1}\\\nonumber
&=\alpha^{(1-i)(n-1)}\prod_{1\leq j\leq n,j\neq i}(1-\alpha^{j-i})^{-1}=\alpha^{(1-i)(n-1)}\prod_{1\leq i\leq n-1}(1-\alpha^{i})^{-1}\\
&=n^{-1}\alpha^{i-1}.
\end{split}
\end{equation}
The last equality follows by the fact $\prod_{1\leq i\leq n-1}(x-\alpha^{i})=\sum_{i=0}^{n-1} x^{i}$.

Consider the $l$-Galois hull of the
$[n,k]_{q}$ MDS code $\mathcal{C}:=GRS_{k}(\mathbf{a},\mathbf{v})$. Then for any
$\mathbf{c}=(f(\alpha^{0}),\ldots,f(\alpha^{n-z}),v_{n-z+2}f(\alpha^{n-z+1}),\ldots,v_{n}f(\alpha^{n-1}))\in Hull_{l}(\mathcal{C})$ with $\mbox {deg}(f(x))\leq k-1$, by Lemma 2 there exists a polynomial
$g(x)\in \mathbb{F}_{q}[x]$ with $\mbox {deg}(g(x))\leq n-k-1$, such that
$$(f^{p^{l}}(\alpha^{0}),\ldots,f^{p^{l}}(\alpha^{n-z}),v_{n-z+2}^{p^{l}+1}f^{p^{l}}(\alpha^{n-z+1}),\ldots,v_{n}^{p^{l}+1}f^{p^{l}}(\alpha^{n-1}))$$
\begin{align}\label{align:1}
=(n^{-1}g(\alpha^{0}),n^{-1}\alpha g(\alpha),\ldots,n^{-1}\alpha^{n-1}g(\alpha^{n-1})).
\end{align}

Comparing the first $n-z+1$ coordinates of Eq. (1), one has
$f^{p^{l}}(\alpha^{i})=n^{-1}\alpha^{i}g(\alpha^{i})$, $i=0,\ldots,n-z$. Hence the number of the distinct roots of $f^{p^{l}}(x)-n^{-1}xg(x)$ is at least
$n-z+1\geq n-k+1$. On the other hand, since $\mbox {deg}(f^{p^{l}}(x))\leq p^{l}(k-1)\leq n-k-1$ and
$\mbox {deg}(xg(x))\leq n-k$, one has $\mbox {deg}(f^{p^{l}}(x)-n^{-1}xg(x))\leq n-k$. Thus $f^{p^{l}}(x)=n^{-1}xg(x)$, which yields
$x\mid f(x)$.

Moreover, comparing the last $z-1$ coordinates of Eq. (1), one has
$$v_{i}^{p^{l}+1}f^{p^{l}}(\alpha^{i-1})=n^{-1}\alpha^{i-1}g(\alpha^{i-1})=f^{p^{l}}(\alpha^{i-1}),$$
namely $f(\alpha^{i-1})=0$ for $i=n-z+2,\ldots,n$. Then we can express $f(x)$ as
$$f(x)=xc(x)\prod_{i=n-z+1}^{n-1}(x-\alpha^{i})$$
for some $c(x)\in \mathbb{F}_{q}[x]$ with $\mbox {deg}(c(x))\leq k-z-1$. Thus $\mbox {dim}(Hull_{l}(\mathcal{C}))\leq k-z$.

Conversely, let $f(x)=xc(x)\prod_{i=n-z+1}^{n-1}(x-\alpha^{i})$,
where $c(x)\in\mathbb{F}_{q}[x]$ with $\mathrm{deg}(c(x))\leq k-z-1$.
Taking $g(x)=nx^{-1}f^{p^{l}}(x)$, then $\mathrm {deg}(g(x))\leq p^{l}(k-1)-1\leq n-k-1$ and
$$(f^{p^{l}}(\alpha^{0}),\ldots,f^{p^{l}}(\alpha^{n-z}),v_{n-z+2}^{p^{l}+1}f^{p^{l}}(\alpha^{n-z+1}),\ldots,v_{n}^{p^{l}+1}f^{p^{l}}(\alpha^{n-1}))$$
\begin{align*}\label{align:1}
=(n^{-1}g(\alpha^{0}),n^{-1}\alpha g(\alpha),\ldots,n^{-1}\alpha^{n-1}g(\alpha^{n-1})).
\end{align*}
By Lemma 2, we have $(f(\alpha^{0}),\ldots,f(\alpha^{n-z}),v_{n-z+2}f(\alpha^{n-z+1}),\ldots,v_{n}f(\alpha^{n-1}))\in Hull_{l}(\mathcal{C})$,
which illustrates that $\mbox {dim}(Hull_{l}(\mathcal{C}))\geq k-z$. Therefore, we have
$\mbox {dim}(Hull_{l}(\mathcal{C}))=k-z=h$.

(2) For $n\mid (p^{l}-1)$, there exists a primitive $n$th root of unity in $\mathbb{F}_{p^{l}}$. We still denote it as $\alpha$ and set $z:=k-h$.
Besides, let $\mathbf{a}$ and $\mathbf{v}$ be defined as in part (1). Write $f(x)=\sum_{i=0}^{k-1} f_{i}x^{i}$ and let
$F(x)=\sum_{i=0}^{k-1} f_{i}^{p^{l}}x^{i}$. Clearly, $f^{p^{l}}(\alpha^{j})=F(\alpha^{j})$ for $j=0,\ldots,n-1$. Substituting this into Eq. (1), we obtain
$$(F(\alpha^{0}),\ldots,F(\alpha^{n-z}),v_{n-z+2}^{p^{l}+1}F(\alpha^{n-z+1}),\ldots,v_{n}^{p^{l}+1}F(\alpha^{n-1}))$$
\begin{align}
=(n^{-1}g(\alpha^{0}),n^{-1}\alpha g(\alpha),\ldots,n^{-1}\alpha^{n-1}g(\alpha^{n-1})).
\end{align}

According to the first $n-z+1$ coordinates of Eq. (2), we have $F(\alpha^{i})=n^{-1}\alpha^{i}g(\alpha^{i})$, $i=0,\ldots,n-z$. Hence the number of the distinct roots of $F(x)-n^{-1}xg(x)$ is at least $n-z+1\geq n-k+1$. Since $k\leq \lfloor\frac{n}{2}\rfloor$, we have $\mbox {deg}(F(x))\leq k-1\leq n-k-1$, which, together with
$\mbox {deg}(xg(x))\leq n-k$ derives that $\mbox {deg}(F(x)-n^{-1}xg(x))\leq n-k$.
We thus have $F(x)=n^{-1}xg(x)$. Then $f(0)=0$ since $F(0)=0$.

By the last $z-1$ coordinates of Eq. (2), one has
$$v_{j}^{p^{l}+1}F(\alpha^{j-1})=n^{-1}\alpha^{j-1}g(\alpha^{j-1})=F(\alpha^{j-1}), j=n-z+2,\ldots,n$$
Hence $F(\alpha^{j-1})=0$, namely $f(\alpha^{j-1})=0$ for $j=n-z+2,\ldots,n$. Then we can express $f(x)$ as
$$f(x)=xc(x)\prod_{j=n-z+1}^{n-1}(x-\alpha^{j})$$
for some $c(x)\in \mathbb{F}_{q}[x]$ with $\mbox {deg}(c(x))\leq k-z-1$. Thus $\mbox {dim}(Hull_{l}(\mathcal{C}))\leq k-z$.

Conversely, let $f(x)=xc(x)\prod_{j=n-z+1}^{n-1}(x-\alpha^{j})$, where $c(x)\in \mathbb{F}_{q}[x]$ with $\mathrm{deg}(c(x))\leq k-z-1$.
Taking $g(x)=nx^{-1}F(x)$, then $\mathrm {deg}(g(x))\leq k-2\leq n-k-2$ and Eq. (1) holds.
By Lemma 2, we have $(f(\alpha^{0}),\ldots,f(\alpha^{n-z}),v_{n-z+2}f(\alpha^{n-z+1}),\ldots,v_{n}f(\alpha^{n-1}))\in Hull_{l}(\mathcal{C})$,
which illustrates that $\mbox {dim}(Hull_{l}(\mathcal{C}))\geq k-z$. Therefore, we have
$\mbox {dim}(Hull_{l}(\mathcal{C}))=k-z=h$. $\hfill\square$
\end{proof}

The following lemma will be used to construct other new families of MDS codes with $l$-Galois hulls of arbitrary dimensions.

\begin{lemma} \label{lemma4}
Let $q=p^{e}$ with $p$ being an odd prime number and let $0\leq l\leq e-1$. Then for any $u\in \mathbb{F}_{p^{l}}^{\ast}$, there exists
$v\in \mathbb{F}_{q}^{\ast}$ such that $v^{p^{l}+1}=u$ if and only if $2l\mid e$.
\end{lemma}

\begin{proof}
Since $\mathbb{F}_{p^{l}}$ is required to be a subfield of $\mathbb{F}_{q}$, we immediately obtain $l\mid e$. Let $e=ll'$ for some integer $l'$. Assume
$\mathbb{F}_{q}^{\ast}=\langle \varepsilon \rangle$, then
$\mathrm {ord}(\varepsilon^{p^{l}+1})=\frac{q-1}{\mathrm {gcd}(q-1,p^{l}+1)}$. Denote
$$H:=\{x^{p^{l}+1}|x\in \mathbb{F}_{q}^{\ast}\},$$
then $H$ is a multiplicative subgroup of $\mathbb{F}_{q}^{\ast}$ of order $\frac{q-1}{\mathrm {gcd}(q-1,p^{l}+1)}$.
Note that $\mathbb{F}_{p^{l}}^{\ast}$ is a multiplicative subgroup of $\mathbb{F}_{q}^{\ast}$ of order $p^{l}-1$.
Therefore,
\begin{equation}
\begin{split}
\mathbb{F}_{p^{l}}^{\ast}\subseteq H&\Leftrightarrow (p^{l}-1)\mid \frac{q-1}{\mbox {gcd}(q-1,p^{l}+1)}\\\nonumber
&\Leftrightarrow (p^{l}-1)\cdot \mbox {gcd}(q-1,p^{l}+1)\mid (q-1).
\end{split}
\end{equation}
Note that
\begin{equation}
\begin{split}
\mbox {gcd}(q-1,p^{l}+1)&=\mbox {gcd}(((p^{l}+1)-1)^{l'}-1,p^{l}+1)\\\nonumber
&=\mbox {gcd}((-1)^{l'}-1,p^{l}+1)\\
&=\begin{cases}
2,&l'\  \mbox  {is  odd},\\
p^{l}+1,&l'\  \mbox  {is  even}.
\end{cases}\\
\end{split}
\end{equation}

(1) If $l'$ is odd, then $(p^{l}-1)\cdot \mbox {gcd}(q-1,p^{l}+1)=2(p^{l}-1)$. On the other hand,
$q-1=p^{ll'}-1=(p^{l}-1)\sum_{i=0}^{l^{'}-1} p^{li}$ and $\sum_{i=0}^{l^{'}-1} p^{li}$ is odd. This yields that
$$(p^{l}-1)\cdot \mbox {gcd}(q-1,p^{l}+1)\nmid (q-1).$$

(2) If $l'$ is even, we may assume $l'=2l''$ for some integer $l''$.
Then $(p^{l}-1)\cdot \mbox {gcd}(q-1,p^{l}+1)=p^{2l}-1$. Since $q-1=p^{ll'}-1=p^{2ll''}-1$, we have
$$(p^{l}-1)\cdot \mbox {gcd}(q-1,p^{l}+1)\mid (q-1).$$

Thus, $\mathbb{F}_{p^{l}}^{\ast}\subseteq H\Leftrightarrow 2l\mid e$. This completes the proof. $\hfill\square$
\end{proof}

By Lemma 4, we obtain the following MDS code $\mathcal{C}$ with $h$-dimensional $l$-Galois hull.

\begin{theorem} \label{Theorem2}
Let $q=p^{e}$ with $p$ being an odd prime number and let $0\leq l\leq e-1$. Let $n\leq p^{l}$ with $2l\mid e$. Then for any $0\leq h\leq k$, where
$1\leq k\leq \lfloor\frac{n}{2}\rfloor$, there exists an $[n,k]_{q}$ MDS code $\mathcal{C}$ with $h$-dimensional $l$-Galois hull.
\end{theorem}

\begin{proof}
Since $n\leq p^{l}$, we can take $n$ distinct elements $a_{1},a_{2},\ldots,a_{n}$ from $\mathbb{F}_{p^{l}}$.
Then $u_{i}\in \mathbb{F}_{p^{l}}^{\ast}$. By Lemma 4, there exists
$v_{i}\in \mathbb{F}_{q}^{\ast}$ such that $v_{i}^{p^{l}+1}=u_{i}$ for $i=1,\ldots,n$.
Set $z:=k-h$ and take $\beta\in \mathbb{F}_{q}^{\ast}$ such that $\gamma:=\beta^{p^{l}+1}\neq 1$.
Put $\mathbf{a}=(a_{1},a_{2},\ldots,a_{n})$ and
$\mathbf{v}=(\beta v_{1},\ldots,\beta v_{z},v_{z+1},\ldots,v_{n})$.
Consider the $l$-Galois hull of the
$[n,k]_{q}$ MDS code $\mathcal{C}:=GRS_{k}(\mathbf{a},\mathbf{v})$. Then for any
$\mathbf{c}=(\beta v_{1}f(a_{1}),\ldots,\beta v_{z}f(a_{z}),v_{z+1}f(a_{z+1}),\ldots,v_{n}f(a_{n}))\in Hull_{l}(\mathcal{C})$ with $\mbox {deg}(f(x))\leq k-1$, by Lemma 2 there exists a polynomial
$g(x)\in \mathbb{F}_{q}[x]$ with $\mbox {deg}(g(x))\leq n-k-1$, such that
$$(\beta^{p^{l}+1}v_{1}^{p^{l}+1}f^{p^{l}}(a_{1}),\ldots,\beta^{p^{l}+1}v_{z}^{p^{l}+1}f^{p^{l}}(a_{z})
,v_{z+1}^{p^{l}+1}f^{p^{l}}(a_{z+1}),\ldots,v_{n}^{p^{l}+1}f^{p^{l}}(a_{n}))$$
\begin{align*}
=(u_{1}g(a_{1}),u_{2}g(a_{2}),\ldots,u_{n}g(a_{n})).
\end{align*}
Namely,
$$(\gamma u_{1}f^{p^{l}}(a_{1}),\ldots,\gamma u_{z}f^{p^{l}}(a_{z})
,u_{z+1}f^{p^{l}}(a_{z+1}),\ldots,u_{n}f^{p^{l}}(a_{n}))$$
\begin{align}
=(u_{1}g(a_{1}),u_{2}g(a_{2}),\ldots,u_{n}g(a_{n})).
\end{align}
Write $f(x)=\sum_{i=0}^{k-1} f_{i}x^{i}$ and let
$F(x)=\sum_{i=0}^{k-1} f_{i}^{p^{l}}x^{i}$. Since $a_{i}\in \mathbb{F}_{p^{l}}$, we have $F(a_{i})=f^{p^{l}}(a_{i})$ for $i=1,\ldots,n$.
Then Eq. (3) is transformed into
$$(\gamma u_{1}F(a_{1}),\ldots,\gamma u_{z}F(a_{z})
,u_{z+1}F(a_{z+1}),\ldots,u_{n}F(a_{n}))$$
\begin{align}\label{align:4}
=(u_{1}g(a_{1}),u_{2}g(a_{2}),\ldots,u_{n}g(a_{n})).
\end{align}

In terms of the last $n-z$ coordinates of Eq. (4), we have $F(a_{i})=g(a_{i})$, $i=z+1,\ldots,n$. Hence the number of the distinct roots of
$F(x)-g(x)$ is at least $n-z\geq n-k$. Since $k\leq \lfloor\frac{n}{2}\rfloor$, we have $\mbox {deg}(F(x))\leq k-1\leq n-k-1$, which, together with
$\mbox {deg}(g(x))\leq n-k-1$ derives that $\mbox {deg}(F(x)-g(x))\leq n-k-1$.
We thus have $F(x)=g(x)$.

By the first $z$ coordinates of Eq. (4), we have
$$\gamma u_{j}F(a_{j})=u_{j}g(a_{j})=u_{j}F(a_{j}), j=1,\ldots,z$$
Hence $F(a_{j})=0$, namely $f(a_{j})=0$ for $j=1,\ldots,z$. Then we can express $f(x)$ as
$$f(x)=c(x)\prod_{j=1}^{z}(x-a_{j})$$
for some $c(x)\in \mathbb{F}_{q}[x]$ with $\mbox {deg}(c(x))\leq k-z-1$. Thus $\mbox {dim}(Hull_{l}(\mathcal{C}))\leq k-z$.

Conversely, let $f(x)=c(x)\prod_{j=1}^{z}(x-a_{j})$, where $c(x)\in\mathbb{F}_{q}[x]$ with $\mbox {deg}(c(x))\leq k-z-1$.
Taking $g(x)=F(x)$, then $\mathrm{deg}(g(x))\leq k-1\leq n-k-1$ and $g(a_{i})=F(a_{i})=f^{p^{l}}(a_{i})$. Hence,
$$(\beta^{p^{l}+1}v_{1}^{p^{l}+1}f^{p^{l}}(a_{1}),\ldots,\beta^{p^{l}+1}v_{z}^{p^{l}+1}f^{p^{l}}(a_{z})
,v_{z+1}^{p^{l}+1}f^{p^{l}}(a_{z+1}),\ldots,v_{n}^{p^{l}+1}f^{p^{l}}(a_{n}))$$
\begin{align*}
=(u_{1}g(a_{1}),u_{2}g(a_{2}),\ldots,u_{n}g(a_{n})).
\end{align*}
By Lemma 2, we have
$$(\beta v_{1}f(a_{1}),\ldots,\beta v_{z}f(a_{z}),v_{z+1}f(a_{z+1}),\ldots,v_{n}f(a_{n}))\in Hull_{l}(\mathcal{C}).$$
Then $\mbox {dim}(Hull_{l}(\mathcal{C}))\geq k-z$. Therefore, $\mbox {dim}(Hull_{l}(\mathcal{C}))=k-z=h$. $\hfill\square$
\end{proof}

Let $\mathrm{ord}(x)$ denote the order of the element $x$ in $\mathbb{F}_{q}^{\ast}$. We have the following result.

\begin{lemma}
Let $\xi_{1}=\alpha^{x_{1}}$ and $\xi_{2}=\alpha^{x_{2}}$, where $\alpha$ is a primitive element of $\mathbb{F}_{q}$. Then
$$\mathrm {gcd}(\mathrm{ord}(\xi_{1}),\mathrm{ord}(\xi_{2}))=1 \Leftrightarrow (q-1)\mid \mathrm{lcm}(x_{1},x_{2}).$$
\end{lemma}

\begin{proof}
Since $\mathrm{ord}(\xi_{1})=\frac{q-1}{\mathrm {gcd}(x_{1},q-1)}$ and
$\mathrm{ord}(\xi_{2})=\frac{q-1}{\mathrm {gcd}(x_{2},q-1)}$, then
$\mathrm {gcd}(\mathrm{ord}(\xi_{1}),\mathrm{ord}(\xi_{2}))=1$ if and only if
\begin{align}
\mathrm {gcd}\bigg(\frac{q-1}{\mathrm {gcd}(x_{1},q-1)},\frac{q-1}{\mathrm {gcd}(x_{2},q-1)}\bigg)=1.
\end{align}
Let $S$ be the set consisting of all the prime divisors of $q-1$, $x_{1}$ and $x_{2}$.
Assume $q-1=\prod_{p_{i}\in S}p_{i}^{\alpha_{i}}$, $x_{1}=\prod_{p_{i}\in S}p_{i}^{\beta_{i}}$,
$x_{2}=\prod_{p_{i}\in S} p_{i}^{\gamma_{i}}$,
where $\alpha_{i},\beta_{i},\gamma_{i}\in \mathbb{N}$. Then
$$\mathrm{gcd}(x_{1},q-1)=\prod_{p_{i}\in S} p_{i}^{\mathrm{min}(\alpha_{i},\beta_{i})},$$
which yields that
$$\frac{q-1}{\mathrm{gcd}(x_{1},q-1)}=\prod_{p_{i}\in S} p_{i}^{\alpha_{i}-\mathrm{min}(\alpha_{i},\beta_{i})}.$$
Hence, Eq. (5) holds if and only if for each $i$,
\begin{align}
\begin{split}
0&= \mathrm{min}(\alpha_{i}-\mathrm{min}(\alpha_{i},\beta_{i}),\alpha_{i}-\mathrm{min}(\alpha_{i},\gamma_{i}))\\\nonumber
&= \alpha_{i}-\mathrm{max}(\mathrm{min}(\alpha_{i},\beta_{i}),\mathrm{min}(\alpha_{i},\gamma_{i})).
\end{split}
\end{align}
That is,
\begin{align*}
  \alpha_{i}=\mathrm{max}(\mathrm{min}(\alpha_{i},\beta_{i}),\mathrm{min}(\alpha_{i},\gamma_{i}))
 &\Leftrightarrow \alpha_{i}=\mathrm{min}(\alpha_{i},\beta_{i}) \     \mbox{or}    \  \alpha_{i}=\mathrm{min}(\alpha_{i},\gamma_{i})  \\
 &\Leftrightarrow \alpha_{i}\leq \beta_{i} \     \mbox{or}    \  \alpha_{i}\leq \gamma_{i} \\
 &\Leftrightarrow \alpha_{i}\leq  \mathrm{max}(\beta_{i},\gamma_{i})  \\
 &\Leftrightarrow \prod_{p_{i}\in S} p_{i}^{\alpha_{i}}\Bigm| \mathrm{lcm}\Big(\prod_{p_{i}\in S}p_{i}^{\beta_{i}},\prod_{p_{i}\in S}p_{i}^{\gamma_{i}}\Big)  \\
 &\Leftrightarrow (q-1)\mid \mathrm{lcm}(x_{1},x_{2}).
\end{align*}
This completes the proof. $\hfill\square$
\end{proof}

\vspace{4pt}
Let $\alpha$ be a primitive element of $\mathbb{F}_{q}$. Consider $\xi_{1}=\alpha^{x_{1}}$ and $\xi_{2}=\alpha^{x_{2}}$.
Let $n=r_{1}r_{2}$, where $1\leq r_{1}\leq \mathrm{ord}(\xi_{1})$, $r_{2}=\mathrm{ord}(\xi_{2})$.
Denote
\begin{align}\label{align:5}
\mathcal{R}=\bigcup_{i=1}^{r_{1}} R_{i}=\{a_{1},\ldots,a_{n}\},
\end{align}
where $R_{i}=\{\xi_{1}^{i}\xi_{2}^{j}|j=1,\ldots,r_{2}\}$ for $i=1,2,\ldots,r_{1}$. Then we have the following result.

\begin{lemma} \label{lemma6}
With the above notations. Assume $l\mid e$, $(q-1)\mid \mathrm{lcm}(x_{1},x_{2})$ and $\mathrm{gcd}(x_{2},q-1)\mid x_{1}(p^{l}-1)$.
For any $i=1,\ldots,n$, let $a_{i}\in R_{s}$ for some $1\leq s\leq r_{1}$. Then $a_{i}=\xi_{1}^{s}\xi_{2}^{t}$
for some $1\leq t\leq r_{2}$. We obtain
$$u_{i}=a_{i}\xi_{1}^{-sr_{2}}r_{2}^{-1} \prod_{1\leq s'\leq r_{1},s'\neq s}(\xi_{1}^{sr_{2}}-\xi_{1}^{s'r_{2}})^{-1}.$$
Besides, $a_{i}^{-1}u_{i}\in \mathbb{F}_{p^{l}}^{\ast}$.
\end{lemma}

\begin{proof}
Since $(q-1)\mid \mathrm{lcm}(x_{1},x_{2})$, by Lemma 5, $\mathrm {gcd}(\mathrm{ord}(\xi_{1}),\mathrm{ord}(\xi_{2}))=1$ holds, which means that
$\langle \xi_{1} \rangle \bigotimes \langle \xi_{2} \rangle$ is a subgroup of $\mathbb{F}_{q}^{\ast}$ with order
$\mbox{ord}(\xi_{1})\cdot \mbox{ord}(\xi_{2})$. Hence, $a_{i}\neq a_{j}$ for any $1\leq i\neq j\leq n$. One can see that
$$u_{i}=\prod_{a_{j}\in R_{s},a_{i}\neq a_{j}}(a_{i}-a_{j})^{-1}\cdot \prod_{1\leq s'\leq r_{1},s'\neq s}\prod_{a_{j'}\in R_{s'}}(a_{i}-a_{j'})^{-1}.$$
Note that $\prod_{1\leq t'\leq r_{2}-1}(x-\xi_{2}^{t'})=\sum_{i=0}^{r_{2}-1} x^{i}$, then
$$\prod_{a_{j}\in R_{s},a_{i}\neq a_{j}}(a_{i}-a_{j})=\prod_{1\leq t'\leq r_{2},t'\neq t} (\xi_{1}^{s}\xi_{2}^{t}-\xi_{1}^{s}\xi_{2}^{t'})=(\xi_{1}^{s}\xi_{2}^{t})^{r_{2}-1}\prod_{1\leq t'\leq r_{2}-1}(1-\xi_{2}^{t'})=a_{i}^{-1}\xi_{1}^{sr_{2}}r_{2}.$$
Besides, in light of $\prod_{1\leq t'\leq r_{2}}(x-b\xi_{2}^{t'})=x^{r_{2}}-b^{r_{2}}$, we have
\begin{equation}
\begin{split}
\prod_{a_{j'}\in R_{s'}}(a_{i}-a_{j'})&=\prod_{1\leq t'\leq r_{2}}(\xi_{1}^{s}\xi_{2}^{t}-\xi_{1}^{s'}\xi_{2}^{t'})=\xi_{1}^{sr_{2}}-\xi_{1}^{s'r_{2}}.\\\nonumber
\end{split}
\end{equation}
Therefore, $u_{i}$ is obtained.

One can check that
$\xi_{1}^{r_{2}}\in \mathbb{F}_{p^{l}}^{\ast}$ if and only if $\mathrm{gcd}(x_{2},q-1)\mid x_{1}(p^{l}-1)$.
Then $a_{i}^{-1}u_{i}\in \mathbb{F}_{p^{l}}^{\ast}$. $\hfill\square$
\end{proof}

\begin{lemma}
Let $q=p^{e},l\mid e$. Then for two positive integers $x_{1}$ and $x_{2}$, the following statements are equivalent:

(1) $(q-1)\mid \mathrm{lcm}(x_{1},x_{2})$, $\mathrm{gcd}(x_{2},q-1)\mid x_{1}(p^{l}-1)$;

(2) $(q-1)\mid \mathrm{lcm}(x_{1},x_{2})$, $\frac{q-1}{p^{l}-1}\mid x_{1}$.
\end{lemma}

\begin{proof}
(2)$\Rightarrow$(1): For $\frac{q-1}{p^{l}-1}\mid x_{1}$, one has $(q-1)\mid x_{1}(p^{l}-1)$, which immediately yields that
$\mathrm{gcd}(x_{2},q-1)\mid x_{1}(p^{l}-1)$.

\vspace{6pt}
(1)$\Rightarrow$(2): Note that for any $a,b,c\in \mathbb{N}$,
$$a\mid \mathrm{lcm}(b,c)\Leftrightarrow a\mid\mathrm{lcm}(b,\mathrm{gcd}(a,c)).$$
Then by $(q-1)\mid \mathrm{lcm}(x_{1},x_{2})$, we have
\begin{align}
(q-1)\mid \mathrm{lcm}(x_{1},\mathrm{gcd}(x_{2},q-1)).
\end{align}
Besides, it follows from $\mathrm{gcd}(x_{2},q-1)\mid x_{1}(p^{l}-1)$ that
\begin{align*}
\mathrm{lcm}(x_{1},\mathrm{gcd}(x_{2},q-1))\mid \mathrm{lcm}(x_{1},x_{1}(p^{l}-1)),
\end{align*}
namely
\begin{align}
\mathrm{lcm}(x_{1},\mathrm{gcd}(x_{2},q-1))\mid x_{1}(p^{l}-1).
\end{align}
Combining Eqs. (7) and (8), we obtain $(q-1)\mid x_{1}(p^{l}-1)$, then $\frac{q-1}{p^{l}-1}\mid x_{1}$.

This completes the proof. $\hfill\square$
\end{proof}

Based on the previous lemmas, we have the following theorem.

\begin{theorem}
Let $q=p^{e}$ with $p$ being an odd prime number. Assume $2l\mid e$, $(q-1)\mid \mathrm{lcm}(x_{1},x_{2})$ and $\frac{q-1}{p^{l}-1}\mid x_{1}$.
Let $n=\frac{r(q-1)}{\mathrm{gcd}(x_{2},q-1)}$, where $1\leq r\leq \frac{q-1}{\mathrm{gcd}(x_{1},q-1)}$. Then for any
$1\leq k\leq \lfloor\frac{p^{l}+n}{p^{l}+1}\rfloor$,

(1) there exists an $[n,k]_{q}$ MDS code $\mathcal{C}$ with $h$-dimensional $l$-Galois hull for any $0\leq h\leq k-1$;

(2) there exists an $[n+1,k]_{q}$ MDS code $\mathcal{C}$ with $h$-dimensional $l$-Galois hull for any $0\leq h\leq k$;

(3) there exists an $[n+2,k]_{q}$ MDS code $\mathcal{C}$ with $h$-dimensional $l$-Galois hull for any $0\leq h\leq k-1$.
\end{theorem}

\begin{proof}
(1) Let $a_{1},a_{2},\ldots,a_{n}$ be defined as in Eq. (6). By Lemmas 4, 6 and 7, there exists
$v_{i}\in \mathbb{F}_{q}^{\ast}$ such that $v_{i}^{p^{l}+1}=a_{i}^{-1}u_{i}$ for $i=1,\ldots,n$.
Set $z:=k-1-h$ and take $\beta\in \mathbb{F}_{q}^{\ast}$ such that $\gamma:=\beta^{p^{l}+1}\neq 1$.
Put $\mathbf{a}=(a_{1},a_{2},\ldots,a_{n})$ and
$\mathbf{v}=(\beta v_{1},\ldots,\beta v_{z},v_{z+1},\ldots,v_{n})$.
Consider the $l$-Galois hull of the
$[n,k]_{q}$ MDS code $\mathcal{C}:=GRS_{k}(\mathbf{a},\mathbf{v})$. Then for any
$\mathbf{c}=(\beta v_{1}f(a_{1}),\ldots,\beta v_{z}f(a_{z}),v_{z+1}f(a_{z+1}),\ldots,v_{n}f(a_{n}))\in Hull_{l}(\mathcal{C})$ with $\mbox {deg}(f(x))\leq k-1$, by Lemma 1 there exists a polynomial
$g(x)\in \mathbb{F}_{q}[x]$ with $\mbox {deg}(g(x))\leq n-k-1$, such that
$$(\beta^{p^{l}+1}v_{1}^{p^{l}+1}f^{p^{l}}(a_{1}),\ldots,\beta^{p^{l}+1}v_{z}^{p^{l}+1}f^{p^{l}}(a_{z})
,v_{z+1}^{p^{l}+1}f^{p^{l}}(a_{z+1}),\ldots,v_{n}^{p^{l}+1}f^{p^{l}}(a_{n}))$$
\begin{align*}
=(u_{1}g(a_{1}),u_{2}g(a_{2}),\ldots,u_{n}g(a_{n})).
\end{align*}
Namely,
$$(\gamma a_{1}^{-1}u_{1}f^{p^{l}}(a_{1}),\ldots,\gamma a_{z}^{-1}u_{z}f^{p^{l}}(a_{z})
,a_{z+1}^{-1}u_{z+1}f^{p^{l}}(a_{z+1}),\ldots,a_{n}^{-1}u_{n}f^{p^{l}}(a_{n}))$$
\begin{align}\label{align:3}
=(u_{1}g(a_{1}),u_{2}g(a_{2}),\ldots,u_{n}g(a_{n})).
\end{align}

Comparing the last $n-z$ coordinates of Eq. (9), we have $a_{i}^{-1}u_{i}f^{p^{l}}(a_{i})=u_{i}g(a_{i})$, i.e. $f^{p^{l}}(a_{i})=a_{i}g(a_{i})$ for $i=z+1,\ldots,n$. Hence the number of the distinct roots of
$f^{p^{l}}(x)-xg(x)$ is at least $n-z\geq n-k+1$. Since $k\leq \lfloor\frac{p^{l}+n}{p^{l}+1}\rfloor$, we have $\mbox {deg}(f^{p^{l}}(x))\leq p^{l}(k-1)\leq n-k$, which, together with
$\mbox {deg}(xg(x))\leq n-k$ derives that $\mbox {deg}(f^{p^{l}}(x)-xg(x))\leq n-k$.
We thus have $f^{p^{l}}(x)=xg(x)$ and $x\mid f(x)$.

Observing the first $z$ coordinates of Eq. (9), we have
$$\gamma a_{i}^{-1}u_{i}f^{p^{l}}(a_{i})=u_{i}g(a_{i})=u_{i}a_{i}^{-1}f^{p^{l}}(a_{i})$$
for $i=1,\ldots,z$. Hence $f^{p^{l}}(a_{i})=0$, i.e. $f(a_{i})=0$ for $i=1,\ldots,z$. Then we can express $f(x)$ as
$$f(x)=xc(x)\prod_{i=1}^{z}(x-a_{i})$$
for some $c(x)\in \mathbb{F}_{q}[x]$ with $\mbox {deg}(c(x))\leq k-z-2$. Thus $\mbox {dim}(Hull_{l}(\mathcal{C}))\leq k-z-1$.

Conversely, similar to the proof of Theorem 1, we have $\mbox {dim}(Hull_{l}(\mathcal{C}))\geq k-z-1$. Therefore,
$\mbox {dim}(Hull_{l}(\mathcal{C}))=k-z-1=h$.

\vspace{4pt}
(2) Let $a_{1},\ldots,a_{n}$ be defined as in Eq. (6) and let $a_{n+1}=0$. By Lemma 6, we have
$$\prod_{1\leq j\leq n+1,j\neq i}(a_{i}-a_{j})^{-1}=a_{i}^{-1}\prod_{1\leq j\leq n,j\neq i}(a_{i}-a_{j})^{-1}\in \mathbb{F}_{p^{l}}^{\ast}$$
for any $1\leq i\leq n$. For $i=n+1$, we have
$$\prod_{j=1}^{n}(a_{n+1}-a_{j})^{-1}=(-1)^{n}\Big[\prod_{j=1}^{r_{2}}(\prod_{i=1}^{r_{1}}\xi_{1}^{i}\xi_{2}^{j})\Big]^{-1}
=(-1)^{n}\xi_{1}^{-\frac{r_{1}(r_{1}+1)r_{2}}{2}}\xi_{2}^{-\frac{r_{2}(r_{2}+1)r_{1}}{2}}\in \mathbb{F}_{p^{l}}^{\ast}.$$
Denote $w_{i}=\prod_{1\leq j\leq n+1,j\neq i}(a_{i}-a_{j})^{-1},i=1,\ldots,n+1$. Then from Lemma 4, there exists
$v_{i}\in \mathbb{F}_{q}^{\ast}$ such that $v_{i}^{p^{l}+1}=w_{i}$ for $i=1,\ldots,n+1$.
Set $z:=k-h$ and take $\beta\in \mathbb{F}_{q}^{\ast}$ such that $\gamma:=\beta^{p^{l}+1}\neq 1$.
Put $\mathbf{a}=(a_{1},a_{2},\ldots,a_{n+1})$ and
$\mathbf{v}=(\beta v_{1},\ldots,\beta v_{z},v_{z+1},\ldots,v_{n+1})$.
Consider the $l$-Galois hull of the
$[n+1,k]_{q}$ MDS code $\mathcal{C}:=GRS_{k}(\mathbf{a},\mathbf{v})$. Then for any
$\mathbf{c}=(\beta v_{1}f(a_{1}),\ldots,\beta v_{z}f(a_{z}),v_{z+1}f(a_{z+1}),\ldots,v_{n+1}f(a_{n+1}))\in Hull_{l}(\mathcal{C})$ with $\mbox {deg}(f(x))\leq k-1$, by Lemma 2 there exists a polynomial
$g(x)\in \mathbb{F}_{q}[x]$ with $\mbox {deg}(g(x))\leq n-k$, such that
$$(\beta^{p^{l}+1}v_{1}^{p^{l}+1}f^{p^{l}}(a_{1}),\ldots,\beta^{p^{l}+1}v_{z}^{p^{l}+1}f^{p^{l}}(a_{z})
,v_{z+1}^{p^{l}+1}f^{p^{l}}(a_{z+1}),\ldots,v_{n+1}^{p^{l}+1}f^{p^{l}}(a_{n+1}))$$
\begin{align*}
=(w_{1}g(a_{1}),w_{2}g(a_{2}),\ldots,w_{n+1}g(a_{n+1})).
\end{align*}
Namely,
$$(\gamma w_{1}f^{p^{l}}(a_{1}),\ldots,\gamma w_{z}f^{p^{l}}(a_{z})
,w_{z+1}f^{p^{l}}(a_{z+1}),\ldots,w_{n+1}f^{p^{l}}(a_{n+1}))$$
\begin{align}\label{align:3}
=(w_{1}g(a_{1}),w_{2}g(a_{2}),\ldots,w_{n+1}g(a_{n+1})).
\end{align}

From the last $n-z+1$ coordinates of Eq. (10), we have $w_{i}f^{p^{l}}(a_{i})=w_{i}g(a_{i})$, i.e. $f^{p^{l}}(a_{i})=g(a_{i})$ for $i=z+1,\ldots,n+1$. Hence the number of the distinct roots of
$f^{p^{l}}(x)-g(x)$ is at least $n-z+1\geq n-k+1$. Since $k\leq \lfloor\frac{p^{l}+n}{p^{l}+1}\rfloor$, we have $\mbox {deg}(f^{p^{l}}(x))\leq p^{l}(k-1)\leq n-k$, which, together with
$\mbox {deg}(g(x))\leq n-k$ derives that $\mbox {deg}(f^{p^{l}}(x)-g(x))\leq n-k$.
Hence $f^{p^{l}}(x)=g(x)$.

Observing the first $z$ coordinates of Eq. (10), we have
$$\gamma w_{i}f^{p^{l}}(a_{i})=w_{i}g(a_{i})=w_{i}f^{p^{l}}(a_{i})$$
for $i=1,\ldots,z$. Hence $f^{p^{l}}(a_{i})=0$, i.e. $f(a_{i})=0$ for $i=1,\ldots,z$. Then we can express $f(x)$ as
$$f(x)=c(x)\prod_{i=1}^{z}(x-a_{i})$$
for some $c(x)\in \mathbb{F}_{q}[x]$ with $\mbox {deg}(c(x))\leq k-z-1$. Thus $\mbox {dim}(Hull_{l}(\mathcal{C}))\leq k-z$.

Conversely, similar to the proof of Theorem 2, we have $\mbox {dim}(Hull_{l}(\mathcal{C}))\geq k-z$.
Therefore, $\mbox {dim}(Hull_{l}(\mathcal{C}))=k-z=h$.
\vspace{4pt}

(3) Let $z:=k-1-h$. Let $\mathbf{a}=(a_{1},a_{2},\ldots,a_{n+1})$, $\mathbf{v}=(\beta v_{1},\ldots,\beta v_{z},v_{z+1},\ldots,v_{n+1})$
and $w_{i}$ be defined as in part (2).
Consider the $l$-Galois hull of the
$[n+2,k]_{q}$ MDS code $\mathcal{C}:=GRS_{k}(\mathbf{a},\mathbf{v},\infty)$. For any
$\mathbf{c}=(\beta v_{1}f(a_{1}),\ldots,\beta v_{z}f(a_{z}),v_{z+1}f(a_{z+1}),\ldots,v_{n+1}f(a_{n+1}),f_{k-1})\in Hull_{l}(\mathcal{C})$ with $\mbox {deg}(f(x))\leq k-1$, by Lemma 3 there exists a polynomial
$g(x)\in \mathbb{F}_{q}[x]$ with $\mbox {deg}(g(x))\leq n-k+1$, such that
$$(\beta^{p^{l}+1}v_{1}^{p^{l}+1}f^{p^{l}}(a_{1}),\ldots,\beta^{p^{l}+1}v_{z}^{p^{l}+1}f^{p^{l}}(a_{z})
,v_{z+1}^{p^{l}+1}f^{p^{l}}(a_{z+1}),\ldots,v_{n+1}^{p^{l}+1}f^{p^{l}}(a_{n+1}),f_{k-1}^{p^{l}})$$
\begin{align*}
=(w_{1}g(a_{1}),\ldots,w_{n+1}g(a_{n+1}),-g_{n-k+1}).
\end{align*}
Namely,
$$(\gamma w_{1}f^{p^{l}}(a_{1}),\ldots,\gamma w_{z}f^{p^{l}}(a_{z})
,w_{z+1}f^{p^{l}}(a_{z+1}),\ldots,w_{n+1}f^{p^{l}}(a_{n+1}),f_{k-1}^{p^{l}})$$
\begin{align}\label{align:8}
=(w_{1}g(a_{1}),\ldots,w_{n+1}g(a_{n+1}),-g_{n-k+1}).
\end{align}

For $i=z+1,\ldots,n+1$, by comparing the $i$-th coordinate of Eq. (11), we have $w_{i}f^{p^{l}}(a_{i})=w_{i}g(a_{i})$, i.e. $f^{p^{l}}(a_{i})=g(a_{i})$. Hence the number of the distinct roots of
$f^{p^{l}}(x)-g(x)$ is at least $n-z+1\geq n-k+2$. Since $k\leq \lfloor\frac{p^{l}+n}{p^{l}+1}\rfloor$, we have $\mbox {deg}(f^{p^{l}}(x))\leq p^{l}(k-1)\leq n-k$, which, together with
$\mbox {deg}(g(x))\leq n-k+1$ derives that $\mbox {deg}(f^{p^{l}}(x)-g(x))\leq n-k+1$.
Hence $f^{p^{l}}(x)=g(x)$.

Moreover, we have $f_{k-1}^{p^{l}}=-g_{n-k+1}$ from Eq. (11). Assume that $f_{k-1}\neq 0$. By
$\mbox {deg}(f^{p^{l}}(x))=\mbox {deg}(g(x))$, we have $p^{l}(k-1)=n-k+1$, which yields a contradiction to $p^{l}(k-1)\leq n-k$.
Hence, $f_{k-1}=0$, implying that $\mbox {deg}(f(x))\leq k-2$.

According to the first $z$ coordinates of Eq. (11), we have
$$\gamma w_{i}f^{p^{l}}(a_{i})=w_{i}g(a_{i})=w_{i}f^{p^{l}}(a_{i})$$
for $i=1,\ldots,z$. Hence $f^{p^{l}}(a_{i})=0$, i.e. $f(a_{i})=0$ for $i=1,\ldots,z$. Then $f(x)$ can be expressed as
$$f(x)=c(x)\prod_{i=1}^{z}(x-a_{i})$$
for some $c(x)\in \mathbb{F}_{q}[x]$ with $\mbox {deg}(c(x))\leq k-2-z$. Thus $\mbox {dim}(Hull_{l}(\mathcal{C}))\leq k-1-z$.

Conversely, similar to the proof of Theorem 2, we have $\mbox {dim}(Hull_{l}(\mathcal{C}))\geq k-1-z$.
Therefore, $\mbox {dim}(Hull_{l}(\mathcal{C}))=k-1-z=h$. $\hfill\square$
\end{proof}

\vspace{4pt}
Let $q=p^{e}$ with $p$ being a prime number. Assume $l\mid e$ and set $y:=\frac{q-1}{p^{l}-1}$. Let $m\mid(q-1)$. Then $m$ can be labeled as $m=m_{1}m_{2}$, where
$m_{1}=\frac{m}{\mathrm{gcd}(m,y)}$ and $m_{2}=\mbox {gcd}(m,y)$.
Let $\mathbb{F}_{q}^{\ast}=\langle \alpha \rangle$. Denote $H=\langle \vartheta_{1} \rangle$ and
$G=\langle \vartheta_{2} \rangle$, where $\vartheta_{1}:=\alpha^{\frac{q-1}{m}}$ and $\vartheta_{2}=\alpha^{\frac{y}{m_{2}}}$.
Then $\mbox {ord}(H)=m$ and $\mbox {ord}(G)=(p^{l}-1)m_{2}$.

Combining $m_{1}\mid \frac{q-1}{m_{2}}$, i.e. $m_{1}\mid (p^{l}-1)\cdot \frac{y}{m_{2}}$ with
$\mbox {gcd}(m_{1},\frac{y}{m_{2}})=1$, we obtain $m_{1}\mid (p^{l}-1)$, which implies that $H$ is a subgroup of $G$.
We may assume $\eta_{i}$ to be coset representatives of $G/H$, where $i=1,\ldots,\frac{p^{l}-1}{m_{1}}$.
Let $n=rm$, where $1\leq r\leq \frac{p^{l}-1}{m_{1}}$. Denote
$$\mathcal{H}=\bigcup_{i=1}^{r} \eta_{i}H=\{a_{1},\ldots,a_{n}\}.$$
We have the following result.

\begin{lemma} \label{lemma6}
With the above notations. For any $i=1,\ldots,n$, let $a_{i}\in \eta_{s}H$ for some $1\leq s\leq r$. Then $a_{i}=\eta_{s}\vartheta_{1}^{t}$
for some $1\leq t\leq m$. We obtain
$$u_{i}=a_{i}\eta_{s}^{-m}m^{-1} \prod_{1\leq s'\leq r,s'\neq s}(\eta_{s}^{m}-\eta_{s'}^{m})^{-1}.$$
Besides, $a_{i}^{-1}u_{i}\in \mathbb{F}_{p^{l}}^{\ast}$.
\end{lemma}

\begin{proof}
One can see that
$$u_{i}=\prod_{a_{j}\in \eta_{s}H,a_{i}\neq a_{j}}(a_{i}-a_{j})^{-1}\cdot \prod_{1\leq s'\leq r,s'\neq s}\prod_{a_{j'}\in \eta_{s'}H}(a_{i}-a_{j'})^{-1}.$$
We have
\begin{equation}
\begin{split}
\prod_{a_{j}\in \eta_{s}H,a_{i}\neq a_{j}}(a_{i}-a_{j})&=\prod_{1\leq t'\leq m,t'\neq t}(\eta_{s}\vartheta_{1}^{t}-\eta_{s}\vartheta_{1}^{t'})=(\eta_{s}\vartheta_{1}^{t})^{m-1}\prod_{1\leq t'\leq m-1}(1-\vartheta_{1}^{t'})=a_{i}^{-1}\eta_{s}^{m}m,\\\nonumber
\end{split}
\end{equation}
and we also obtain
\begin{equation}
\begin{split}
\prod_{a_{j'}\in \eta_{s'}H}(a_{i}-a_{j'})=\prod_{1\leq t'\leq m}(\eta_{s}\vartheta_{1}^{t}-\eta_{s'}\vartheta_{1}^{t'})=\eta_{s}^{m}-\eta_{s'}^{m}.\\\nonumber
\end{split}
\end{equation}
Then $u_{i}$ can be calculated.

For any $1\leq i\leq r$, $\eta_{i}=\vartheta_{2}^{j}$ for some $1\leq j\leq (p^{l}-1)m_{2}$. Then $\eta_{i}^{m}=\alpha^{jm_{1}y}\in\mathbb{F}_{p^{l}}^{\ast}$, which means that $a_{i}^{-1}u_{i}\in \mathbb{F}_{p^{l}}^{\ast}$.  $\hfill\square$
\end{proof}

By Lemma 6, we have the following theorem.
\begin{theorem} \label{Theorem5}
Let $q=p^{e}$ with $p$ being an odd prime number. Assume $2l\mid e$ and $m\mid (q-1)$. Set $y:=\frac{q-1}{p^{l}-1}$.
Let $n=rm$, where $1\leq r\leq \frac{p^{l}-1}{m_{1}}$ with $m_{1}=\frac{m}{\mathrm{gcd}(m,y)}$.
Then for any
$1\leq k\leq \lfloor\frac{p^{l}+n}{p^{l}+1}\rfloor$,

(1) there exists an $[n,k]_{q}$ MDS code $\mathcal{C}$ with $h$-dimensional $l$-Galois hull for any $0\leq h\leq k-1$;

(2) there exists an $[n+1,k]_{q}$ MDS code $\mathcal{C}$ with $h$-dimensional $l$-Galois hull for any $0\leq h\leq k$;

(3) there exists an $[n+2,k]_{q}$ MDS code $\mathcal{C}$ with $h$-dimensional $l$-Galois hull for any $0\leq h\leq k-1$.
\end{theorem}

\begin{proof}
Similar to the proof of Theorem 3. $\hfill\square$
\end{proof}

\begin{remark} \label{Remark3}
Note that when $e=2l$, Theorem 4 is the same to the results in \cite[Theorems 3.8-3.10]{Fang2019Euclidean}. Therefore, Theorem 4
is a generalization of \cite[Theorems 3.8-3.10]{Fang2019Euclidean}.
\end{remark}

\begin{remark}
As far as we know, the MDS codes with Galois hulls of arbitrary dimension in Theorems 1-4 are constructed for the first time.
\end{remark}

\section{Construction of EAQECCs}
In this section, we will construct nine families of EAQECCs from the previous results in Section 3.
First, let's review some basic concepts and notations on quantum codes.

For the complex field $\mathbb{C}$, let $\mathbb{C}^{q}$ denote the $q$-dimensional Hilbert space over $\mathbb{C}$.
For a pure $n$-qudit state, it can be represented as
$|\mathbf{v}\rangle=\sum_{\mathbf{a}\in \mathbb{F}_{q}^{n}} v_{\mathbf{a}} |\mathbf{a}\rangle$, where $v_{\mathbf{a}}\in \mathbb{C}$
with $\sum_{\mathbf{a}\in \mathbb{F}_{q}^{n}} |v_{\mathbf{a}}|^{2}=1$ and
$\{|\mathbf{a}\rangle=|a_{1}\rangle \bigotimes \cdots \bigotimes|a_{n}\rangle:(a_{1},\ldots,a_{n})\in \mathbb{F}_{q}^{n}\}$
is a basis of $\mathbb{C}^{q^{n}}$. For $\mathbf{a}=(a_{1},\ldots,a_{n})$, $\mathbf{b}=(b_{1},\ldots,b_{n})\in \mathbb{F}_{q}^{n}$,
let $T(\mathbf{a})=T(a_{1})\bigotimes\cdots\bigotimes T(a_{n})$ and
$R(\mathbf{a})=R(a_{1})\bigotimes\cdots\bigotimes R(a_{n})$ be the tensor products of $n$ error operators. Then the error set
$$E_{n}=\{\gamma^{i}T(\mathbf{a})R(\mathbf{b})|0\leq i\leq p-1,\mathbf{a},\mathbf{b}\in \mathbb{F}_{q}^{n}\}$$
forms an error group, where $\gamma$ is a complex primitive $p$-th root of unity, $T(\mathbf{a})$ and $R(\mathbf{b})$ are defined as
$T(\mathbf{a})|\mathbf{v}\rangle=|\mathbf{v}+\mathbf{a}\rangle$ and
$R(\mathbf{b})|\mathbf{v}\rangle=\gamma^{tr_{\mathbb{F}_{q}/\mathbb{F}_{p}}((\mathbf{b},\mathbf{v})_E)}|\mathbf{v}\rangle$, respectively.
For any error $\mathbf{e}=\gamma^{i}T(\mathbf{a})R(\mathbf{b})$, its quantum weight is defined by
$w_{Q}(\mathbf{e})=\sharp\{i|(a_{i},b_{i})\neq (0,0)\}$.
Let $E_{n}(i)=\{\mathbf{e}\in E_{n}|w_{Q}(\mathbf{e})\leq i\}$. For a $q$-ary quantum code $Q$, if $d$
is the largest integer such that $\langle \mathbf{u}|\mathbf{e}|\mathbf{v}\rangle=\mathbf{0}$ for any
$|\mathbf{u}\rangle,|\mathbf{v}\rangle\in Q$ with
$\langle \mathbf{u}|\mathbf{v}\rangle=\mathbf{0}$ and
$\mathbf{e}\in E_{n}(d-1)$, then $Q$ has minimum distance $d$.

We denote by $[[n,k,d]]_{q}$ a $q$-ary quantum code of length $n$, dimension $q^{k}$ and minimum distance $d$.
It has the abilities to detect $d-1$ quantum errors and correct $\lfloor \frac{d-1}{2}\rfloor$ quantum errors.
Moreover, the code rate $\frac{k}{n}$ of such a quantum code is also an important parameter in practice. For fixed
length $n$ and minimum distance $d$, the larger value of $\frac{k}{n}$ (or $k$) means the better performance of the quantum code.

In the theory of quantum error correction, a significant development was the introduction of entanglement-assisted quantum error-correcting codes (EAQECCs).
It is assumed that in addition to a quantum channel, the sender and the receiver share a certain amount of pre-existing entangled bits.
By utilizing the shared entangled bits, it is possible that the sender can
send more qubits or correct more errors in the sense of the same rate of transmission.
We denote by $[[n,k,d;c]]_{q}$ a $q$-ary EAQECC which encodes $k$ logical qubits into $n$ channel qubits
by means of $c$ entangled bits. For an EAQECC, its performance is determined by its rate
$\frac{k}{n}$ and net rate $\frac{k-c}{n}$. The value of net rate can be positive,
negative or zero. When $c=0$, the EAQECC is just a quantum stabilizer code. The parameters of an EAQECC satify the following quantum
Singleton bound.

\begin{lemma}
(\cite{Brun2006Correcting}) The parameters of any $[[n,k,d;c]]_{q}$ EAQECC satisfy
$$n+c-k\geq 2(d-1),$$
where $0\leq c\leq n-1$.
\end{lemma}

If the parameters of an EAQECC satisfy the quantum Singleton bound, i.e. $n+c-k= 2(d-1)$,
the it is called a MDS EAQECC.

The following result provided by Wilde and Brun \cite{Wilde2008Optimal} tells us that an EAQECC can be constructed from two classical codes.

\begin{lemma}
(\cite{Wilde2008Optimal}) Assume that $\mathcal{C}_{1}:[n,k_{1},d_{1}]_{q}$ and $\mathcal{C}_{2}:[n,k_{2},d_{2}]_{q}$ are two
linear codes with parity check matrices $H_{1}$ and $H_{2}$, respectively. Then there exists an
$[[n,k_{1}+k_{2}-n+c,\mathrm{min}\{d_{1},d_{2}\};c]]_{q}$ EAQECC, where $c=\mathrm{rank}(H_{1}H_{2}^{T})$
is the required number of maximally entangled states.
\end{lemma}

For a matrix $A=(a_{ij})$ over $\mathbb{F}_{q}$, define $A^{(p^{e-l})}=(a_{ij}^{p^{e-l}})$ and denote
$A^{\ddag}=[A^{(p^{e-l})}]^{T}$. Then we have the following lemma.

\begin{lemma}
(\cite{Liu2020New}) Let $\mathcal{C}$ be a linear $[n,k,d]_{q}$ code with parity-check matrix $H$. Then
$$\mathrm{rank}(HH^{\ddag})=n-k-\mathrm{dim}(Hull_{l}(\mathcal{C})).$$
\end{lemma}

Taking $\mathcal{C}_{1}=\mathcal{C}$ and $\mathcal{C}_{2}=\mathcal{C}^{(p^{e-l})}$ in Lemma 10 and applying Lemma 11, 
one can obtain the following result.

\begin{lemma}
Let $\mathcal{C}$ be a linear $[n,k,d]_{q}$ code with parity-check matrix $H$. Then
there exists an $[[n,k-\mathrm{dim}(Hull_{l}(\mathcal{C})),d;n-k-\mathrm{dim}(Hull_{l}(\mathcal{C}))]]_{q}$ EAQECC. Further,
if $\mathcal{C}$ is an $[n,k]_{q}$ MDS code, then exists an $[[n,k-\mathrm{dim}(Hull_{l}(\mathcal{C})),n-k+1;n-k-\mathrm{dim}(Hull_{l}(\mathcal{C}))]]_{q}$
MDS EAQECC.
\end{lemma}

Applying Theorems 1-4 to Lemma 12, we construct the following nine families of MDS EAQECCs with flexible parameters.
\begin{theorem}

For any $0\leq h\leq k$, there exists an $[[n,k-h,n-k+1;n-k-h]]_{q}$ MDS EAQECC if one of the following conditions holds:

(1) $n\mid (q-1)$, $1\leq k\leq \lfloor\frac{p^{l}+n-1}{p^{l}+1}\rfloor$;

(2) $n\mid (p^{l}-1)$, $1\leq k\leq \lfloor\frac{n}{2}\rfloor$.

\end{theorem}

\begin{theorem}
Let $q=p^{e}$ with $p$ being an odd prime number and let $0\leq l\leq e-1$. Let $n\leq p^{l}$ with $2l\mid e$. Then for any $0\leq h\leq k$, where
$1\leq k\leq \lfloor\frac{n}{2}\rfloor$, there exists an $[[n,k-h,n-k+1;n-k-h]]_{q}$ MDS EAQECC.
\end{theorem}

\begin{theorem}
Let $q=p^{e}$ with $p$ being an odd prime number. Assume $2l\mid e$, $(q-1)\mid \mathrm{lcm}(x_{1},x_{2})$ and $\frac{q-1}{p^{l}-1}\mid x_{1}$.
Let $n=\frac{r(q-1)}{\mathrm{gcd}(x_{2},q-1)}$, where $1\leq r\leq \frac{q-1}{\mathrm{gcd}(x_{1},q-1)}$. Then for any
$1\leq k\leq \lfloor\frac{p^{l}+n}{p^{l}+1}\rfloor$,

(1) there exists an $[[n,k-h,n-k+1;n-k-h]]_{q}$ MDS EAQECC for any $0\leq h\leq k-1$;

(2) there exists an $[[n+1,k-h,n-k+2;n-k-h+1]]_{q}$ MDS EAQECC for any $0\leq h\leq k$;

(3) there exists an $[[n+2,k-h,n-k+3;n-k-h+2]]_{q}$ MDS EAQECC for any $0\leq h\leq k-1$.
\end{theorem}

\begin{theorem} \label{Theorem5}
Let $q=p^{e}$ with $p$ being an odd prime number. Assume $2l\mid e$ and $m\mid (q-1)$. Set $y:=\frac{q-1}{p^{l}-1}$.
Let $n=rm$, where $1\leq r\leq \frac{p^{l}-1}{m_{1}}$ with $m_{1}=\frac{m}{\mathrm{gcd}(m,y)}$.
Then for any
$1\leq k\leq \lfloor\frac{p^{l}+n}{p^{l}+1}\rfloor$,

(1) there exists an $[[n,k-h,n-k+1;n-k-h]]_{q}$ MDS EAQECC for any $0\leq h\leq k-1$;

(2) there exists an $[[n+1,k-h,n-k+2;n-k-h+1]]_{q}$ MDS EAQECC for any $0\leq h\leq k$;

(3) there exists an $[[n+2,k-h,n-k+3;n-k-h+2]]_{q}$ MDS EAQECC for any $0\leq h\leq k-1$.
\end{theorem}

\begin{remark}
Note that the required number of maximally entangled
states of the MDS EAQECCs shown in \cite{Chen2017Entanglement-assisted,Guenda2018Constructions,Lu2018Entanglement-assisted,Qian2018On} are fixed. By contrast, the required number
of maximally entangled states of the MDS EAQECCs in Theorems 5-8 can take almost all possible values.
Therefore, the MDS EAQECCs in Theorems 5-8 have more flexible parameters.
To illustrate our results in the above theorems, we provide some new MDS EAQECCs in Tables 1-4, which are not obtained by the Euclidean
case and the Hermitian case since $l\neq 0$ and $l\neq\frac{e}{2}$ for even $e$.
\end{remark}

\begin{table}
\centering	
\small
\setlength{\abovecaptionskip}{0.cm}
\setlength{\belowcaptionskip}{0.2cm}
\caption{Some new MDS EAQECCs from Theorem 5 for $e=6$ and $l=2$.}
\begin{tabular}{llllll}
\hline
$k$&$h$&MDS EAQECCs&$k$&$h$&MDS EAQECCs\\
\hline
$10$&$1$&$[[63,9,54;52]]_{64}$&$12$&$3$&$[[63,9,52;48]]_{64}$ \\
$10$&$2$&$[[63,8,54;51]]_{64}$&$12$&$4$&$[[63,8,52;47]]_{64}$ \\
$10$&$3$&$[[63,7,54;50]]_{64}$&$12$&$5$&$[[63,7,52;46]]_{64}$ \\
$10$&$4$&$[[63,6,54;49]]_{64}$&$12$&$6$&$[[63,6,52;45]]_{64}$ \\
$10$&$5$&$[[63,5,54;48]]_{64}$&$12$&$7$&$[[63,5,52;44]]_{64}$ \\
$10$&$6$&$[[63,4,54;47]]_{64}$&$12$&$8$&$[[63,4,52;43]]_{64}$ \\
$10$&$7$&$[[63,3,54;46]]_{64}$&$12$&$9$&$[[63,3,52;42]]_{64}$ \\
$10$&$8$&$[[63,2,54;45]]_{64}$&$12$&$10$&$[[63,2,52;41]]_{64}$ \\
$10$&$9$&$[[63,1,54;44]]_{64}$&$12$&$11$&$[[63,1,52;40]]_{64}$ \\
$10$&$10$&$[[63,0,54;43]]_{64}$&$12$&$12$&$[[63,0,52;39]]_{64}$ \\
$11$&$1$&$[[63,10,53;51]]_{64}$&$13$&$1$&$[[63,12,51;49]]_{64}$ \\
$11$&$2$&$[[63,9,53;50]]_{64}$&$13$&$2$&$[[63,11,51;48]]_{64}$ \\
$11$&$3$&$[[63,8,53;49]]_{64}$&$13$&$3$&$[[63,10,51;47]]_{64}$ \\
$11$&$4$&$[[63,7,53;48]]_{64}$&$13$&$4$&$[[63,9,51;46]]_{64}$ \\
$11$&$5$&$[[63,6,53;47]]_{64}$&$13$&$5$&$[[63,8,51;45]]_{64}$ \\
$11$&$6$&$[[63,5,53;46]]_{64}$&$13$&$6$&$[[63,7,51;44]]_{64}$ \\
$11$&$7$&$[[63,4,53;45]]_{64}$&$13$&$7$&$[[63,6,51;43]]_{64}$ \\
$11$&$8$&$[[63,3,53;44]]_{64}$&$13$&$8$&$[[63,5,51;42]]_{64}$ \\
$11$&$9$&$[[63,2,53;43]]_{64}$&$13$&$9$&$[[63,4,51;41]]_{64}$ \\
$11$&$10$&$[[63,1,53;42]]_{64}$&$13$&$10$&$[[63,3,51;40]]_{64}$ \\
$11$&$11$&$[[63,0,53;41]]_{64}$&$13$&$11$&$[[63,2,51;39]]_{64}$ \\
$12$&$1$&$[[63,11,52;50]]_{64}$&$13$&$12$&$[[63,1,51;38]]_{64}$ \\
$12$&$2$&$[[63,10,52;49]]_{64}$&$13$&$13$&$[[63,0,51;37]]_{64}$ \\
\hline
\end{tabular}
\end{table}

\begin{table}[H]
\centering	
\small
\setlength{\abovecaptionskip}{0.cm}
\setlength{\belowcaptionskip}{0.2cm}
\caption{Some new MDS EAQECCs from Theorem 6 for $e=8$ and $l=2$.}
\begin{tabular}{llllll}
\hline
$k$&$h$&MDS EAQECCs&$k$&$h$&MDS EAQECCs\\
\hline
$9$&$1$&$[[25,8,17;15]]_{5^{8}}$&$11$&$3$&$[[25,8,15;11]]_{5^{8}}$ \\
$9$&$2$&$[[25,7,17;14]]_{5^{8}}$&$11$&$4$&$[[25,7,15;10]]_{5^{8}}$ \\
$9$&$3$&$[[25,6,17;13]]_{5^{8}}$&$11$&$5$&$[[25,6,15;9]]_{5^{8}}$ \\
$9$&$4$&$[[25,5,17;12]]_{5^{8}}$&$11$&$6$&$[[25,5,15;8]]_{5^{8}}$ \\
$9$&$5$&$[[25,4,17;11]]_{5^{8}}$&$11$&$7$&$[[25,4,15;7]]_{5^{8}}$ \\
$9$&$6$&$[[25,3,17;10]]_{5^{8}}$&$11$&$8$&$[[25,3,15;6]]_{5^{8}}$ \\
$9$&$7$&$[[25,2,17;9]]_{5^{8}}$&$11$&$9$&$[[25,2,15;5]]_{5^{8}}$ \\
$9$&$8$&$[[25,1,17;8]]_{5^{8}}$&$11$&$10$&$[[25,1,15;4]]_{5^{8}}$ \\
$9$&$9$&$[[25,0,17;7]]_{5^{8}}$&$11$&$11$&$[[25,0,15;3]]_{5^{8}}$ \\
$10$&$1$&$[[25,9,16;14]]_{5^{8}}$&$12$&$1$&$[[25,11,14;12]]_{5^{8}}$ \\
$10$&$2$&$[[25,8,16;13]]_{5^{8}}$&$12$&$2$&$[[25,10,14;11]]_{5^{8}}$ \\
$10$&$3$&$[[25,7,16;12]]_{5^{8}}$&$12$&$3$&$[[25,9,14;10]]_{5^{8}}$ \\
$10$&$4$&$[[25,6,16;11]]_{5^{8}}$&$12$&$4$&$[[25,8,14;9]]_{5^{8}}$ \\
$10$&$5$&$[[25,5,16;10]]_{5^{8}}$&$12$&$5$&$[[25,7,14;8]]_{5^{8}}$ \\
$10$&$6$&$[[25,4,16;9]]_{5^{8}}$&$12$&$6$&$[[25,6,14;7]]_{5^{8}}$ \\
$10$&$7$&$[[25,3,16;8]]_{5^{8}}$&$12$&$7$&$[[25,5,14;6]]_{5^{8}}$ \\
$10$&$8$&$[[25,2,16;7]]_{5^{8}}$&$12$&$8$&$[[25,4,14;5]]_{5^{8}}$ \\
$10$&$9$&$[[25,1,16;6]]_{5^{8}}$&$12$&$9$&$[[25,3,14;4]]_{5^{8}}$ \\
$10$&$10$&$[[25,0,16;5]]_{5^{8}}$&$12$&$10$&$[[25,2,14;3]]_{5^{8}}$ \\
$11$&$1$&$[[25,10,15;13]]_{5^{8}}$&$12$&$11$&$[[25,1,14;2]]_{5^{8}}$ \\
$11$&$2$&$[[25,9,15;12]]_{5^{8}}$&$12$&$12$&$[[25,0,14;1]]_{5^{8}}$ \\
\hline
\end{tabular}
\end{table}

\begin{table}[H]
\centering	
\small
\setlength{\abovecaptionskip}{0.cm}
\setlength{\belowcaptionskip}{0.2cm}
\caption{Some new MDS EAQECCs from Theorem 7 for $e=4,l=1,x_{1}=160$ and $x_{2}=3$.}
\begin{tabular}{llllll}
\hline
$k$&$h$&MDS EAQECCs&$k$&$h$&MDS EAQECCs\\
\hline
$20$&$6$&$[[80,14,61;54]]_{81}$&$20$&$13$&$[[80,7,61;47]]_{81}$ \\
$20$&$6$&$[[81,14,62;55]]_{81}$&$20$&$13$&$[[81,7,62;48]]_{81}$ \\
$20$&$6$&$[[82,14,63;56]]_{81}$&$20$&$13$&$[[82,7,62;49]]_{81}$ \\
$20$&$7$&$[[80,13,61;53]]_{81}$&$20$&$14$&$[[80,6,61;46]]_{81}$ \\
$20$&$7$&$[[81,13,62;54]]_{81}$&$20$&$14$&$[[81,6,62;47]]_{81}$ \\
$20$&$7$&$[[82,13,63;55]]_{81}$&$20$&$14$&$[[82,6,63;48]]_{81}$ \\
$20$&$8$&$[[80,12,61;52]]_{81}$&$20$&$15$&$[[80,5,61;45]]_{81}$ \\
$20$&$8$&$[[81,12,62;53]]_{81}$&$20$&$15$&$[[81,5,62;46]]_{81}$ \\
$20$&$8$&$[[82,12,63;54]]_{81}$&$20$&$15$&$[[82,5,63;47]]_{81}$ \\
$20$&$9$&$[[80,11,61;51]]_{81}$&$20$&$16$&$[[80,4,61;44]]_{81}$ \\
$20$&$9$&$[[81,11,62;52]]_{81}$&$20$&$16$&$[[81,4,62;45]]_{81}$ \\
$20$&$9$&$[[82,11,63;53]]_{81}$&$20$&$16$&$[[82,4,63;46]]_{81}$ \\
$20$&$10$&$[[80,10,61;50]]_{81}$&$20$&$17$&$[[80,3,61;43]]_{81}$ \\
$20$&$10$&$[[81,10,62;51]]_{81}$&$20$&$17$&$[[81,3,62;44]]_{81}$ \\
$20$&$10$&$[[82,10,63;52]]_{81}$&$20$&$17$&$[[82,3,63;45]]_{81}$ \\
$20$&$11$&$[[80,9,61;49]]_{81}$&$20$&$18$&$[[80,2,61;42]]_{81}$ \\
$20$&$11$&$[[81,9,62;50]]_{81}$&$20$&$18$&$[[81,2,62;43]]_{81}$ \\
$20$&$11$&$[[82,9,63;51]]_{81}$&$20$&$18$&$[[82,2,63;44]]_{81}$ \\
$20$&$12$&$[[80,8,61;48]]_{81}$&$20$&$19$&$[[80,1,61;41]]_{81}$ \\
$20$&$12$&$[[81,8,62;49]]_{81}$&$20$&$19$&$[[81,1,62;42]]_{81}$ \\
$20$&$12$&$[[82,8,63;50]]_{81}$&$20$&$19$&$[[82,1,63;43]]_{81}$ \\
\hline
\end{tabular}
\end{table}

\begin{table}[H]
\centering	
\small
\setlength{\abovecaptionskip}{0.cm}
\setlength{\belowcaptionskip}{0.2cm}
\caption{Some new MDS EAQECCs from Theorem 8 for $e=4$ and $l=1$.}
\begin{tabular}{llllll}
\hline
$k$&$h$&MDS EAQECCs&$k$&$h$&MDS EAQECCs\\
\hline
$9$&$1$&$[[40,8,32;30]]_{81}$&$10$&$1$&$[[40,9,31;29]]_{81}$ \\
$9$&$1$&$[[41,8,33;31]]_{81}$&$10$&$1$&$[[41,9,32;30]]_{81}$ \\
$9$&$1$&$[[42,8,34;32]]_{81}$&$10$&$1$&$[[42,9,33;31]]_{81}$ \\
$9$&$2$&$[[40,7,32;29]]_{81}$&$10$&$2$&$[[40,8,31;28]]_{81}$ \\
$9$&$2$&$[[41,7,33;30]]_{81}$&$10$&$2$&$[[41,8,32;29]]_{81}$ \\
$9$&$2$&$[[42,7,34;31]]_{81}$&$10$&$2$&$[[42,8,33;30]]_{81}$ \\
$9$&$3$&$[[40,6,32;28]]_{81}$&$10$&$3$&$[[40,7,31;27]]_{81}$ \\
$9$&$3$&$[[41,6,33;29]]_{81}$&$10$&$3$&$[[41,7,32;28]]_{81}$ \\
$9$&$3$&$[[42,6,34;30]]_{81}$&$10$&$3$&$[[42,7,33;29]]_{81}$ \\
$9$&$4$&$[[40,5,32;27]]_{81}$&$10$&$4$&$[[40,6,31;26]]_{81}$ \\
$9$&$4$&$[[41,5,33;28]]_{81}$&$10$&$4$&$[[41,6,32;27]]_{81}$ \\
$9$&$4$&$[[42,5,34;29]]_{81}$&$10$&$4$&$[[42,6,33;28]]_{81}$ \\
$9$&$5$&$[[40,4,32;26]]_{81}$&$10$&$5$&$[[40,5,31;25]]_{81}$ \\
$9$&$5$&$[[41,4,33;27]]_{81}$&$10$&$5$&$[[41,5,32;26]]_{81}$ \\
$9$&$5$&$[[42,4,34;28]]_{81}$&$10$&$5$&$[[42,5,33;27]]_{81}$ \\
$9$&$6$&$[[40,3,32;25]]_{81}$&$10$&$6$&$[[40,4,31;24]]_{81}$ \\
$9$&$6$&$[[41,3,33;26]]_{81}$&$10$&$6$&$[[41,4,32;25]]_{81}$ \\
$9$&$6$&$[[42,3,34;27]]_{81}$&$10$&$6$&$[[42,4,33;26]]_{81}$ \\
$9$&$7$&$[[40,2,32;24]]_{81}$&$10$&$7$&$[[40,3,31;23]]_{81}$ \\
$9$&$7$&$[[41,2,33;25]]_{81}$&$10$&$7$&$[[41,3,32;24]]_{81}$ \\
$9$&$7$&$[[42,2,34;26]]_{81}$&$10$&$7$&$[[42,3,33;25]]_{81}$ \\
\hline
\end{tabular}
\end{table}

\section{Conclusion and discussion}

In this paper, by using GRS codes
and extended GRS codes, we provided several new families
of MDS codes with Galois hulls of arbitrary dimensions that are not obtained before.
Some of the GRS codes and extended GRS codes were constructed by virtue of certain
multiplicative subgroups of $\mathbb{F}_{q}^{\ast}$ and their cosets. Moreover, some of them generalized the ones in \cite{Fang2019Euclidean}.
Through these MDS codes, we provided nine new families of MDS EAQECCs in which the required number of maximally entangled states are flexible.
As an illustration, we list some new MDS EAQECCs in Tables 1-4.

Note that the dimension $k$ of the MDS codes constructed in Theorem 1 (1) and Theorems 3 and 4 is roughly bounded by
$\lfloor\frac{p^{l}+n}{p^{l}+1}\rfloor$. Therefore, we propose the following question.

\begin{question}

Can we construct MDS codes with a larger dimension $k$ such that
$\lfloor\frac{p^{l}+n}{p^{l}+1}\rfloor < k\leq \lfloor\frac{n}{2}\rfloor$ and determine the corresponding dimensions of their Galois hulls?

\end{question}

Let $\mathcal{C}$ be an $[n,k]_{q}$ MDS code. By \cite[Proposition 2.4]{Liu2020New}, its $l$-Galois dual code is an $[n,n-k]_{q}$ MDS code.
Then by Lemma 12, there exists an
\begin{equation}
[[n,n-k-\mathrm{dim}(Hull_{l}(\mathcal{C}^{\perp_{l}})),k+1;k-\mathrm{dim}(Hull_{l}(\mathcal{C}^{\perp_{l}}))]]_{q}
\end{equation}
MDS EAQECC. For the Euclidean case, i.e. $l=0$ and for the Hermitian case, i.e. $l=\frac{e}{2}$, since
$(\mathcal{C}^{\perp})^{\perp}=\mathcal{C}$ and $(\mathcal{C}^{\perp_{H}})^{\perp_{H}}=\mathcal{C}$, we have
$$Hull(\mathcal{C})=Hull(\mathcal{C}^{\perp}),      \ \\\   Hull_{H}(\mathcal{C})=Hull_{H}(\mathcal{C}^{\perp_{H}}).$$
As a consequence, the parameters in Eq. (12) can be determined by the dimension of the Euclidean hull or the Hermitian
hull of $\mathcal{C}$. However, for $l\neq0$ and $l\neq\frac{e}{2}$, in general, we have $(\mathcal{C}^{\perp_{l}})^{\perp_{l}}\neq \mathcal{C}$.
Naturally, we propose the following question.

\begin{question}

How to determine the relationship between $Hull_{l}(\mathcal{C})$ and $Hull_{l}(\mathcal{C}^{\perp_{l}})$?
Further, is there an equality that represents the relationship between
$\mathrm{dim}(Hull_{l}(\mathcal{C}))$ and $\mathrm{dim}(Hull_{l}(\mathcal{C}^{\perp_{l}}))$?

\end{question}

If we could solve this problem, another nine families of MDS EAQECCs will be constructed from Eq. (12) except those shown in Theorems 5-8.

\bibliographystyle{plain}
\bibliography{2020.4.4.ref}

\begin{thebibliography}{10}

\bibitem{Assmus1993Designs}
Jr.E.F. Assmus and J.D. Key.
\newblock Designs and their codes.
\newblock {\em Cambridge Tracts in Mathematics, Cambridge, U.K.: Cambridge
  Univ. Press}, 103, 1993.

\bibitem{Bringer2014Orthogonal}
J.~Bringer, C.~Carlet, H.~Chabanne, S.~Guilley, and H.~Maghrebi.
\newblock Orthogonal direct sum masking-a smartcard friendly computation
  paradigm in a code, with builtin protection against side-channel and fault
  attacks.
\newblock {\em In: WISTP, Heraklion, Springer, LNCS}, 8501:40--56, 2014.

\bibitem{Brun2006Correcting}
T.~Brun, I.~Devetak, and M.H. Hsieh.
\newblock Correcting quantum errors with entanglement.
\newblock {\em Science}, 314:436--439, 2006.

\bibitem{Calderbank1998Quantum}
A.R. Calderbank, E.M. Rains, P.W. Shor, and N.J.A. Sloane.
\newblock Quantum error correction via codes over {GF}(4).
\newblock {\em IEEE Trans. Inf. Theory}, 44(4):1369--1387, 1998.

\bibitem{Carlet2014Complementary}
C.~Carlet and S.~Guilley.
\newblock Complementary dual codes for countermeasures to side-channel attacks.
\newblock {\em in Coding Theory and Applications (CIM Series in Mathematical
  Sciences), E. R. Pinto et al., Eds. Cham, Switzerland: Springer-Verlag}, 3,
  2014.

\bibitem{Carlet2018Euclidean}
C.~Carlet, S.~Mesnager, C.~Tang, and Y.~Qi.
\newblock {Euclidean} and {Hermitian LCD MDS} codes.
\newblock {\em Des. Codes Cryptogr.}, 86(11):2605--2618, 2018.

\bibitem{Carlet2018New}
C.~Carlet, S.~Mesnager, C.~Tang, and Y.~Qi.
\newblock New characterization and parametrization of {LCD} codes.
\newblock {\em IEEE Trans. Inf. Theory}, 65(1):39--49, 2018.

\bibitem{Carlet2018Linear}
C.~Carlet, S.~Mesnager, C.~Tang, Y.~Qi, and R.~Pellikaan.
\newblock Linear codes over $\mathbb{F}_{q}$ are equivalent to {LCD} codes for
  $q>3$.
\newblock {\em IEEE Trans. Inf. Theory}, 64(4):3010--3017, 2018.

\bibitem{Chen2018New}
B.~Chen and H.~Liu.
\newblock New constructions of {MDS} codes with complementary duals.
\newblock {\em IEEE Trans. Inf. Theory}, 64(8):5776--5782, 2018.

\bibitem{Chen2017Entanglement-assisted}
J.~Chen, Y.~Huang, C.~Feng, and R.~Chen.
\newblock Entanglement-assisted quantum {MDS} codes constructed from negacyclic
  codes.
\newblock {\em Quantum Inf. Process.}, 16(303), 2017.

\bibitem{Fan2017Galois}
Y.~Fan and L.~Zhang.
\newblock Galois self-dual constacyclic codes.
\newblock {\em Des. Codes Cryptogr.}, 84:473--492, 2017.

\bibitem{Fang2018Two}
W.~Fang and F.~Fu.
\newblock Two new classes of quantum {MDS} codes.
\newblock {\em Finite Fields Appl.}, 53:85--98, 2018.

\bibitem{Fang2019Euclidean}
W.~Fang, F.~Fu, L.~Li, and S.~Zhu.
\newblock Euclidean and {Hermitian} hulls of {MDS} codes and their applications
  to {EAQECC}s.
\newblock {\em IEEE Trans. Inf. Theory (Early Access)}, 2019.

\bibitem{Galindo2018New}
C.~Galindo, O.~Geil, F.~Hernando, and D.~Ruano.
\newblock New binary and ternary {LCD} codes.
\newblock {\em IEEE Trans. Inf. Theory}, 65(2):1008--1016, 2018.

\bibitem{Guenda2018Constructions}
K.~Guenda, S.~Jitman, and T.A. Gulliver.
\newblock Constructions of good entanglement assisted quanutm error correcting
  codes.
\newblock {\em Des. Codes Cryptogr.}, 86:121--136, 2018.

\bibitem{Jin2014A}
L.~Jin.
\newblock A construction of new quantum {MDS} codes.
\newblock {\em IEEE Trans. Inf. Theory}, 60(5):2921--2925, 2014.

\bibitem{Jin2017Construction}
L.~Jin.
\newblock Construction of {MDS} codes with complementary duals.
\newblock {\em IEEE Trans. Inf. Theory}, 63(5):2843 -- 2847, 2017.

\bibitem{Liu2015LCD}
X.~Liu and H.~Liu.
\newblock Lcd codes over finite chain rings.
\newblock {\em Finite Fields Appl.}, 34:1--19, 2015.

\bibitem{Liu2020New}
X.~Liu, H.~Liu, and L.~Yu.
\newblock New {EAQEC} codes constructed from {Galois LCD} codes.
\newblock {\em Quantum Inf. Process.}, 19(20), 2020.

\bibitem{Lu2018Entanglement-assisted}
L.~Lu, R.~Li, L.~Guo, Y.~Ma, and Y.~Liu.
\newblock Entanglement-assisted quantum {MDS} codes from negacyclic codes.
\newblock {\em Quantum Inf. Process.}, 17(69), 2018.

\bibitem{Luo2019MDS}
G.~Luo, X.~Cao, and X.~Chen.
\newblock {MDS} codes with hulls of arbitrary dimensions and their quantum
  error correction.
\newblock {\em IEEE Trans. Inf. Theory}, 65(5):2944--2952, 2019.

\bibitem{Massey1992Linear}
J.L. Massey.
\newblock Linear codes with complementary duals.
\newblock {\em Discrete Math.}, 106-107:337--342, 1992.

\bibitem{Qian2018On}
J.~Qian and L.~Zhang.
\newblock On {MDS} linear complementary dual codes and entanglement-assisted
  quantum codes.
\newblock {\em Des. Codes Cryptogr.}, 86(7):1565--1572, 2018.

\bibitem{Shor1995Scheme}
P.W. Shor.
\newblock Scheme for reducing decoherence in quantum computer memory.
\newblock {\em Phys. Rev. A}, 52(4):R2493, 1995.

\bibitem{Sok2020On}
L.~Sok.
\newblock On {Hermitian LCD} codes and their {Gray} image.
\newblock {\em Finite Fields Appl.}, 62, 2020.

\bibitem{Steane1996Multiple}
A.M. Steane.
\newblock Multiple-particle interference and quantum error correction.
\newblock {\em Proc. R. Soc. London A}, 452:2551--2577, 1996.

\bibitem{Wilde2008Optimal}
M.M. Wilde and T.A. Brun.
\newblock Optimal entanglement formulas for entanglement-assisted quantum
  coding.
\newblock {\em Phys. Rev. A}, 77(064302), 2008.

\end{thebibliography}

\end{document}